\documentclass[10pt,journal,compsoc]{IEEEtran}
\usepackage{comment}

\ifCLASSOPTIONcompsoc
  \usepackage[nocompress]{cite}
\else
  \usepackage{cite}
\fi
\ifCLASSINFOpdf
   \usepackage[pdftex]{graphicx}
   \graphicspath{{../fig/}{../pdf/}{../jpeg/}}
  \DeclareGraphicsExtensions{.svg,.pdf,.jpeg,.png}
\else
  \usepackage[dvips]{graphicx}
  \graphicspath{{../eps/}}
  \DeclareGraphicsExtensions{.eps}
\fi
\usepackage{hyperref}
\makeatletter
\def\UrlAlphabet{%
      \do\a\do\b\do\c\do\d\do\e\do\f\do\g\do\h\do\i\do\j%
      \do\k\do\l\do\m\do\n\do\o\do\p\do\q\do\r\do\s\do\t%
      \do\u\do\v\do\w\do\x\do\y\do\z\do\A\do\B\do\C\do\D%
      \do\E\do\F\do\G\do\H\do\I\do\J\do\K\do\L\do\M\do\N%
      \do\O\do\P\do\Q\do\R\do\S\do\T\do\U\do\V\do\W\do\X%
      \do\Y\do\Z}
\def\UrlDigits{\do\1\do\2\do\3\do\4\do\5\do\6\do\7\do\8\do\9\do\0}
\g@addto@macro{\UrlBreaks}{\UrlOrds}
\g@addto@macro{\UrlBreaks}{\UrlAlphabet}
\g@addto@macro{\UrlBreaks}{\UrlDigits}
\makeatother

\usepackage{multirow}
\usepackage{algorithmic}
\usepackage{algorithm}
\usepackage{amsfonts}
\usepackage[dvipsnames]{xcolor}

\newcommand{\pj}[1]{\textcolor{blue}{Peijie[#1]}}
\newcommand{\shortname}{\emph{NESCL}}
\newcommand{\fullname}{\emph{Neighborhood-Enhanced Supervised Contrastive Loss~(\emph{NESCL})}}

\usepackage{amsmath}
\newcommand\numberthis{\addtocounter{equation}{1}\tag{\theequation}}
\usepackage[caption=true,font=footnotesize]{subfig}

\begin{document}
%
\title{Neighborhood-Enhanced Supervised Contrastive Learning for Collaborative Filtering}
%
%
%
%

\author{Peijie~Sun,~
        Le~Wu,~
        Kun~Zhang,~
        Xiangzhi~Chen,~
        and~Meng~Wang
\IEEEcompsocitemizethanks{
\IEEEcompsocthanksitem P.~Sun is with the Department of Computer Science and Technology, Tsinghua University, Beijing 100084, China.  
\protect Email: sun.hfut@gmail.com.
\IEEEcompsocthanksitem L.~Wu, K. Zhang, X.~Chen, M.~Wang are with the School of Computer and Information, Hefei University of Technology,
Hefei, Anhui 230009, China.
\protect Emails: \{lewu.ustc, zhang1028kun, cxz.hfut, eric.mengwang\}@gmail.com.
}
}

\markboth{Journal of \LaTeX\ Class Files,~Vol.~14, No.~8, November~2022}%
{Shell \MakeLowercase{\textit{et al.}}: Bare Demo of IEEEtran.cls for Computer Society Journals}
%

\IEEEtitleabstractindextext{%
\begin{abstract}
While effective in recommendation tasks, collaborative filtering (CF) techniques face the challenge of data sparsity. Researchers have begun leveraging contrastive learning to introduce additional self-supervised signals to address this. However, this approach often unintentionally distances the target user/item from their collaborative neighbors, limiting its efficacy. In response, we propose a solution that treats the collaborative neighbors of the anchor node as positive samples within the final objective loss function.
This paper focuses on developing two unique supervised contrastive loss functions that effectively combine supervision signals with contrastive loss. We analyze our proposed loss functions through the gradient lens, demonstrating that different positive samples simultaneously influence updating the anchor node's embeddings. These samples' impact depends on their similarities to the anchor node and the negative samples. Using the graph-based collaborative filtering model as our backbone and following the same data augmentation methods as the existing contrastive learning model SGL, we effectively enhance the performance of the recommendation model.
Our proposed \fullname~ model substitutes the contrastive loss function in SGL with our novel loss function, showing marked performance improvement. On three real-world datasets, Yelp2018, Gowalla, and Amazon-Book, our model surpasses the original SGL by 10.09\%, 7.09\%, and 35.36\% on NDCG@20, respectively.
\end{abstract}

}

\maketitle

\IEEEdisplaynontitleabstractindextext

\IEEEpeerreviewmaketitle

\IEEEraisesectionheading{\section{Introduction}\label{sec:introduction}}


\IEEEPARstart{D}{ue} to the information overload issue, recommender models have been widely used in many online platforms, such as Yelp\footnote{https://www.yelp.com/}, Gowalla\footnote{https://go.gowalla.com/}, and Amazon\footnote{https://www.amazon.com/}. 
The recommendation models' main idea is that users with a similar consumed history may have similar preferences, which is also the key idea of the Collaborative Filtering(CF) methods. 
There are two kinds of CF methods, memory-based~\cite{icdm2011SLIM, tois2004ItemKNN, ec2000user-based} and model-based~\cite{uai2012BPR, sigir2020LightGCN, kdd2008SVD++}. According to the research trend in recent years, model-based CF methods have attracted a lot of attention because of their efficient performance. 
Nonetheless, the CF models mainly suffer from the data sparsity issue. 
As the main research direction is how to boost the performance of CF models by improving the effectiveness of the user and item representations, many models are proposed to mine more information to enhance the representations of the users and items~\cite{sigir2019NGCF, aaai2020LRGCCF, sigir2020LightGCN, tkde2022HIN, tkde2022accury_neural_recom, feng2021cross}. 
For example, the SVD++ model is proposed to enhance the model-based methods with the nearest neighbors of the items which are achieved by the ItemKNN method~\cite{kdd2008SVD++}. And LightGCN can utilize higher-order collaborative signals to enhance the representations of users and items~\cite{sigir2020LightGCN}. 

\begin{figure}
\centering
\includegraphics[width=0.5\textwidth]{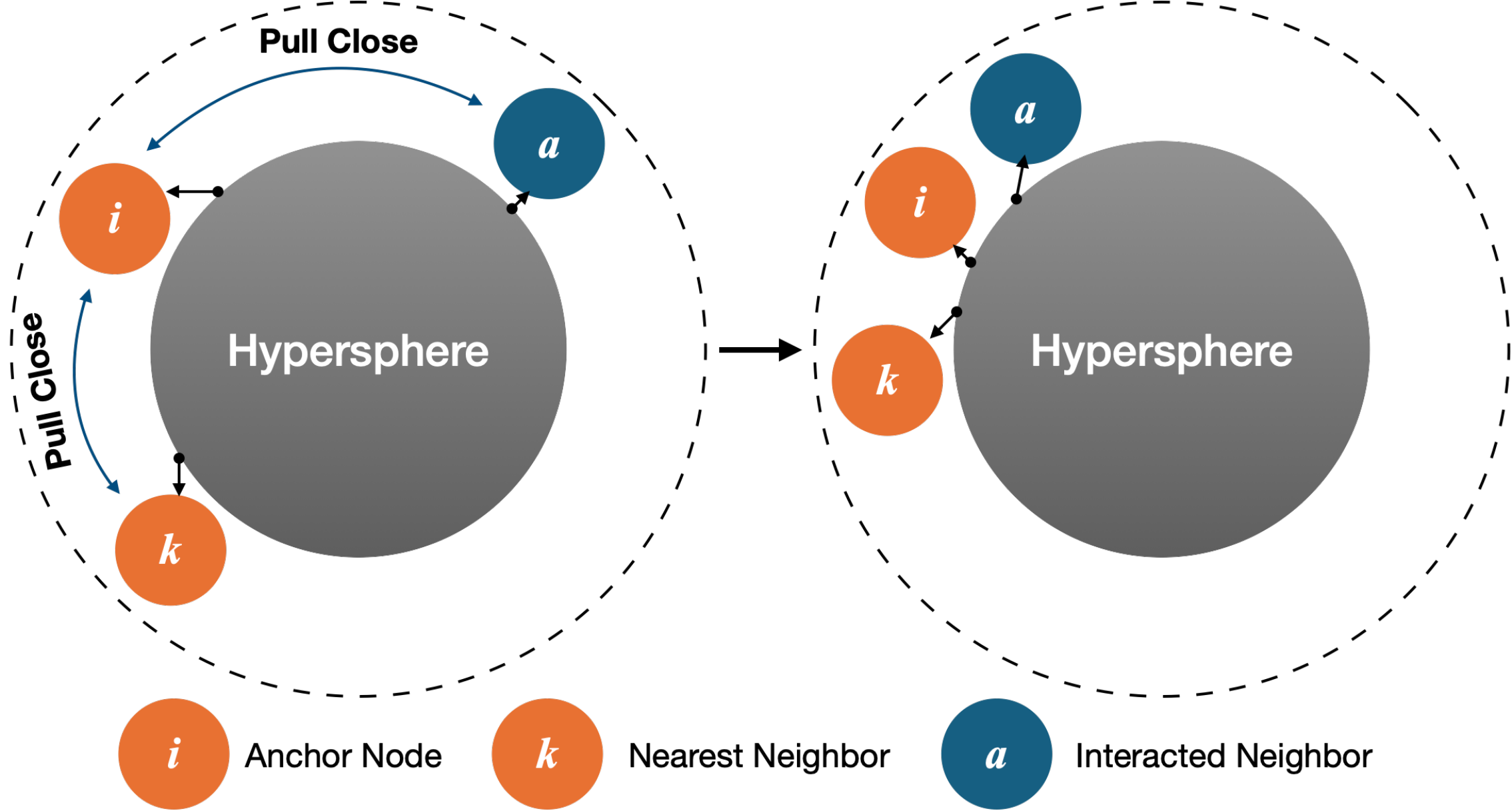}
\caption{We random select an item $i$ as the anchor node. The node $k$ is $i$'s nearest neighbor, which is found by the ItemKNN algorithm, and node $a$ has interacted with item $i$'.}
\label{fig:case}
\end{figure}

Recently, contrastive learning has achieved great success in computer vision areas~\cite{neurips2021SupCon, icml2020SimCLR}. 
As it can provide an additional self-supervised signal, some researchers have tried to introduce it into the recommendation tasks to alleviate the data sparsity issue~\cite{sigir2021SGL, sigir2022GACL, TOIS2022ContraRec, kdd2022DirectAU}. 
The main idea of the contrasitve method is to push apart the anchor node from any other nodes in the representation space. Generally, any user and item can be considered anchor nodes. Other nodes here refer to other users or items.
In recommendation tasks, the representations of the users and items are learned based on their historical interactions. It is a natural idea to generate the augmented data by perturbing the anchor node's historical interaction records.
In the model training stage, the anchor node's representation and its augmented representation are positive samples of each other. Then, other nodes' representations are treated as negative samples. 
 
However, while contrastive learning has shown effectiveness in recommendation tasks, it brings new challenges by potentially distancing anchor nodes from their collaborative neighbors. 
Consequently, some potentially interest-aligned neighbors of the user may be treated as false negative samples in the contrastive loss, undermining the optimization of the recommendation model.
For example, in Figure~\ref{fig:case}, for the anchor node item $i$, the item $k$ and user $a$ are its nearest and interacted neighbors, respectively. The representations of the anchor node and its nearest neighbors and interacted neighbors should be close to each other in the hypersphere. The nearest and interacted neighbors are the anchor node's collaborative neighbors. 
If the contrastive loss optimizes the recommendation model, it will cause the anchor node $i$ to be far away from the collaborative neighbors, such as the left part in Figure~\ref{fig:case}. 
To our best knowledge, few studies have been conducted to address such an issue. In the paper SGL~\cite{sigir2021SGL}, the researchers directly utilize the ranking-based loss function to pull close the anchor node and its interacted neighbors. 
And in the paper ~\cite{www2022NCL}, the authors of NCL studied how to find the positive samples of the anchor node based on the cluster method.

Despite numerous strategies proposed to address the challenging task of integrating supervisory signals with contrastive loss, it remains an intricate problem. We propose a potential solution: treating the collaborative neighbors of the anchor node as positive samples in the final objective loss function. This approach aims to optimize the positioning of all nodes' learned representations in the representation space such that anchor nodes and positive sample nodes are proximate while maximizing the distance from negative sample nodes. Drawing inspiration from the SupCon work~\cite{neurips2021SupCon}, we have devised two novel supervised contrastive loss functions for recommendation tasks. These functions have been meticulously designed to guide the backbone model's optimization more effectively, specifically by focusing on the numerator and denominator of the InfoNCE loss.

In the experimental section, we demonstrate the superior performance of LightGCN, the selected backbone model, trained using our proposed loss functions. Evaluated on three real-world datasets—Yelp2018, Gowalla, and Amazon-Book—our model outshines the current state-of-the-art contrastive learning method, SGL, outperforming it by 10.09\%, 7.09\%, and 35.36\% on the NDCG@20 metric, respectively. We also observe that our method shows enhanced utility with smaller temperature values, indicating that the role of negative samples is amplified at lower temperatures. Despite the presence of false negative samples, using some of the user's nearest neighbors as positive samples for contrastive learning enables us to leverage the advantages of smaller temperature coefficients, thereby offsetting the potential adverse impact of these false samples. Recognizing the potential inaccuracies of algorithm-identified nearest neighbors, we propose strategies to integrate these nearest-neighbor users, enhancing the robustness and performance of our model. This approach provides a novel perspective on managing the variability in the quality of positive samples, promising to pave the way for future advancements in the field.

The contributions of our proposed model can be summarized as follows:

1. We propose an effective model that leverages multiple positive samples of anchor nodes to guide the update of anchor representations. Theoretical analysis shows that the anchor and multiple negative samples jointly determine the influence of different positive samples.

2. Through experiments, we found that our proposed method performs better with a smaller temperature value. This observation reinforces our hypothesis that by introducing multiple positive samples, we can counteract the detrimental effects of false negative samples and amplify the beneficial effects of true negative samples. This work thus provides new insights into optimizing the performance of contrastive loss by adjusting the temperature value.

3. Given the diversity of positive sample types and their limited quality, we propose several strategies for positive sample selection and evaluate the effectiveness of these strategies. Furthermore, our proposed loss function can naturally accommodate various positive sample types, enhancing the model's performance.

Following, we first introduce the work which is related to our work. Then, to help the readers understand the loss functions we proposed, we introduce preliminary knowledge. Next, we will briefly introduce our proposed loss functions and analyze how they work from a theoretical perspective. Last, we conducted experiments on three real-world datasets, to analyze the performance of our proposed model from many perspectives. 
\section{Related Work}

\subsection{Graph Neural Network based Recommender System}
This section focuses on works that use graph neural networks in recommendation tasks. 
CF based models have been widely used for recommending items to users. Among all collaborative filtering based models, latent factor models perform better than other models~\cite{uai2012BPR}. However, the performance of such models is limited because of the data sparsity issue. 
Since the interactions between users and items can be thought of as a user-item bipartite graph, it makes sense that each user's or item's preferences will be affected not only by their first-order neighbors but also by their higher-order neighbors~\cite{dong2021equivalence, feng2021should, sigir2019NGCF, sigir2020LightGCN}. NGCF is proposed to model such higher-order collaborative filtering signal with the help of Graph Neural Network(GCN) technique~\cite{sigir2019NGCF}. 
However, as the original user-item interaction matrix is very sparse, the performance of NGCF is also limited because of its heavy parameters and non-linear activation function for each message-passing layer.
LightGCN is proposed to address the issue of NGCF, by removing the transform parameters and non-linear activation function of NGCF~\cite{sigir2020LightGCN}. As directly utilizing the GCN technique in recommendation tasks may encounter an over-smoothing issue, LRGCCF~\cite{aaai2020LRGCCF} is proposed to alleviate the issue by concatenating the output representations of all users and items among all message-passing layers. 
Even though the LightGCN model has shown surprising performance in recommendation tasks, it is inefficient because of the multiple message-passing layers. The authors in UltraGCN proposed a model named UltraGCN to approximate the stacking message passing operation of LightGCN with a contrastive loss~\cite{cikm2021UltraGCN}. It can reduce the inference time of LightGCN, while it is also time-consuming in the training stage as it has to sample a lot of negative samples.
Besides modeling the user-item bipartite graph, the graph neural network technique is also utilized in other kinds of recommendation tasks, for example, social recommendation~\cite{sigir2019DiffNet, tkde2021DiffNet++}, fraud detection~\cite{www2021PC-GNN}, review-based recommendation~\cite{shuai2022review} and attribute inference~\cite{sigir2020joint}. 

In summary, graph convolutional networks have shown great promise in recommendation tasks, particularly in addressing data sparsity issues and encapsulating higher-order collaborative filtering signals. However, these methods also have shortcomings, such as causing over-smoothing problems. How to alleviate these problems and discover more valuable supervisory signals are still under research. 

\subsection{Self-Supervised Learning Technique}
Self-supervised learning technique has been widely studied in computer vision~\cite{neurips2020BYOL, neurips2021SwAV, cvpr2020MoCo, icml2020SimCLR}, 
natural language processing~\cite{emnlp2021SimCSE, neurips2021R-Drop, acl2021ConSERT}, 
and data mining areas~\cite{iclr2019DGI, neurips2020GraphCL, kdd2020GCC, icml2021JOAO, neurips2021AD-GCL}. 
There are two branches of the self-supervised learning technique, generative~\cite{noroozi2016unsupervised, gidaris2018unsupervised} and contrastive~\cite{icml2020SimCLR, neurips2021SwAV, acl2021ConSERT}. The key idea of the generative self-supervised papers is designing how to reconstruct the corrupted data or predict the input's missing data. And the key idea of the contrastive self-supervised learning technique is how to pull the two augmented representations of the same anchor node close, and push the anchor node's representations away from other nodes' representations.
This paper mainly studies applying the self-supervised technique in user-item bipartite graphs.
However, the self-supervised techniques which are used in the computer vision and natural language processing areas can not be directly used in the graph data because the structure of the graph data is complex and irregular. Current works mainly studied how to design pretext tasks that require the model to make predictions or solve auxiliary tasks based on the input graph~\cite{kdd2020GCC, iclr2019DGI, neurips2020GraphCL}, such as graph reconstruction, node attribution prediction, and so on. 
Due to the sparsity of the user-item rating matrix in the recommendation task, researchers also attempted to use contrastive learning techniques to augment the input data to improve the performance of recommendation tasks such as sequential recommendation~\cite{kdd2020Disseq2seq, TOIS2022ContraRec}, session-based recommendation~\cite{cikm2021COTREC}, social recommendation~\cite{www2021MHCN}, review-based recommendation~\cite{sigir2022RGCL},  and candidate matching tasks~\cite{sigir2021SGL, cikm2021UltraGCN, kdd2022DirectAU}. 

In our paper, we mainly focus on the candidate-matching task. 
In current works, the researchers studied how to augment the representations of the users and items, and how to design contrastive loss to learn a more robust recommendation model~\cite{sigir2021SGL, www2022NCL, sigir2022GACL}. 
One of the key techniques of utilizing contrastive loss is how to augment the data. 
In SGL, the authors in SGL proposed three kinds of data augmentations strategies, node dropout, edge dropout, and random walk to augment the original bipartite graph~\cite{sigir2021SGL}. 
However, in the paper~\cite{arXiv2022sampledsoftmax}, the authors even found that the simple sampled softmax loss itself is capable of mining hard negative samples to enhance the performance of the recommendation models without data augmentation. 
As the data augmentation of SGL~\cite{sigir2021SGL} is time-consuming, the authors in GACL proposed a simple but effective method to augment the representations of the users and items~\cite{sigir2022GACL}, i.e., adding perturbing noise to the representations of the users and items. Some researchers also studied how to utilize positive samples~\cite{www2022NCL}. In this paper NCL~\cite{www2022NCL}, the authors cluster the users and items into several clusters, respectively. And for each anchor node, its corresponding cluster is treated as the positive sample. 
In the paper
~\cite{icml2021Whitening}, the authors proposed a whitening-based method to avoid the representations collapse issue, and they argued that the negative samples are not necessary in the model training stage.
Besides, some researchers studied how to find the negative samples~\cite{arxiv2022NegativeSampling, iclr2021hardness}. 





As our main task is how to design the contrastive loss function to constrain the distance between the anchor node and its collaborative neighbors, to our best knowledge, we found the following two papers, which are related to our work. In the paper~\cite{cvpr2021uniformiy}, the authors further studied the uniformity characteristic of the contrastive loss. They proposed that high uniformity would lead to low tolerance, and make the learned model may push away two samples with similar semantics. Besides, the authors in the paper ~\cite{NeurIPS2022bias} studied how to utilize the popularity degree information to help the collaborative model automatically adjust the collaborative representations optimization intensity of any user-item pair~\cite{NeurIPS2022bias}.
The most related work to ours is SupCon~\cite{neurips2021SupCon}. Though we both proposed two kinds of supervised contrastive loss functions, optimizing the loss functions we proposed model can achieve better performance than the ones in the SupCon. The main difference is the design of the numerator and denominator of the InfoNCE loss. The experimental results also show that training the model based on our proposed loss functions could achieve better than training the model based on the SupCon loss. We suppose that compared to the loss function proposed in Supcon, our proposed loss function is more effective in adaptively tuning the weights of all positive samples. In the section on experiments, the experiments also test how well our proposed loss functions work.

\subsection{Neighbor-based Collaborative Filtering Methods}
As we utilize the neighbor-based methods to find nearest-neighbors of the anchor node, we would simply introduce the neighbor-based collaborative filtering methods. Following, we will simply introduce several kinds of methods that find the nearest neighbors. Then, we will introduce how to utilize the nearest neighbors to help recommendation task.

We split the current nearest-neighbors finding methods into the following three categories. First is finding nearest neighbors based on historical records, such as ItemKNN~\cite{tois2004ItemKNN}, and UserKNN~\cite{ec2000user-based}. Second, to find the nearest neighbors of the cold-start users or items, the researchers incorporate more kinds of data, such as text~\cite{ijcnlp2019NRMS, www2020DualPC}, KG~\cite{www2018DKN}, and social network~\cite{sigir2019DiffNet}, ~\cite{tkde2021DiffNet++}. Third, the researchers aim to find the nearest neighbors with the learned embeddings, such as cluster, and most of current works. 
After getting the nearest neighbors, the data can be used to serve recommendations directly, such as ItemCF~\cite{tois2004ItemKNN} and SLIM~\cite{icdm2011SLIM}, or enhancing the representations of the items, such as SVD++~\cite{kdd2008SVD++}, Diffnet~\cite{sigir2019DiffNet}, and so on. 
\section{Problem Definition \& Preliminary}
\label{sec:preliminary}
In this section, we briefly introduce the preliminary knowledge which is related to our proposed model. First, we introduce the key technique of the backbone model LightGCN~\cite{sigir2020LightGCN} we use. Then, we introduce how augment the input data and how to achieve the augmented users' and items' representations. 

\subsection{Notation \& Problem Definition}
In this paper, we aim to study how to model different kinds of positive samples of the anchor node when designing the supervised contrastive loss. As the backbone model is the LightGCN~\cite{sigir2020LightGCN}, we would introduce the data which is used in the training stage. 
Given the user set $\mathcal{U}(|\mathcal{{U}}|=m)$, item set $\mathcal{V}(|\mathcal{V}|=n)$, and the corresponding rating matrix $\mathbf{R}$, where $\mathbf{R}_{ai}=1$ denotes the user $a$ and item $i$ has interaction, we first construct the bipartite user-item graph $\mathcal{G}=(\mathcal{N}, \mathcal{E})$, where $\mathcal{N}=\mathcal{U}\cup\mathcal{V}$, and $\mathcal{E}$ consists of the connected user-item pairs in the rating matrix $\mathbf{R}$. For any node $i\in\mathcal{N}$, $\mathbf{R}^+_i$ denotes the node $i$'s interacted neighbors. 
Because we treat the anchor nodes' interacted neighbors and nearest neighbors both as collaborative neighbors, we then introduce how to achieve the nearest neighbors simply. 
For example, for any anchor node $i\in\mathcal{N}$, we use $\mathcal{S}_i$ to denote its nearest neighbor set. The number $K$ of the nearest neighbors set $\mathcal{S}_i$ is a hyper-parameter, which is predefined in advance. 
As our proposed loss function is based on the contrastive loss technique, the details about how to augment the input graph and how to get the augmented representations can refer to the following section. 
The important notations which appear in this paper can refer to Table~\ref{tab:notation}.

\begin{table}[htb]
\centering
\caption{Notation Table.}\label{tab:notation}
\small{
\begin{tabular}{c|c}
\hline
Notation                 & Description  \\ \hline
$\mathcal{G}$ &  User-item bipartite graph  \\ \hline
$\mathcal{N}$ &  The union of the user set $\mathcal{U}$ and item item $\mathcal{V}$ \\ \hline
$\mathcal{E}$ & Edge set of the graph $\mathcal{G}$ \\ \hline
$\mathbf{R}^+_i$ & Node $i$'s interacted neighbors \\ \hline
$\mathcal{S}_i$ & Node $i$'s nearest neighbors \\ \hline
$\mathcal{G}'$, $\mathcal{G}''$  & The augmented two graphs \\ \hline
$\mathbf{H}'$, $\mathbf{H}''$  & The representations of all nodes from two views \\ \hline
$\tau$ & The temperature value in the contrastive loss \\ \hline
$\rho$ & The drop ratio in the data augmentation process \\ \hline
\end{tabular}
}
\end{table}

\subsection{Model-based Method: LightGCN}
The main idea of the LightGCN model is modeling the user-item high-order collaborative signal through the GCN network. Given the user-item bipartite graph $\mathcal{G}=(\mathcal{N}, \mathcal{E})$, the LightGCN model can learn the users' and items' representations through $K$ iteration layers. At the $k$-th layer, the learned users' and items' representations $\mathbf{H}^k$ can be treated as containing their $k$-hop neighbors' information. To alleviate the over-smoothing issue in GNN-based models~\cite{aaai2020LRGCCF},   
the representations of all nodes among all propagation layers are concatenated with:
\begin{equation}
    \mathbf{H}=[\mathbf{H}^0, \mathbf{H}^1,...,\mathbf{H}^K].
\end{equation}
The concatenated representations ${\mathbf{H}}$ are also treated as the final representations of all users and items. For user $a$ and item $i$, their representations are denoted as $\mathbf{h}_a$ and $\mathbf{h}_i$, respectively.
Then, for any pair of user-item $(a, i)$, their predicted rating $\hat{r}_{ai}$ can be calculated with the inner product operation:
\begin{equation}
    \hat{r}_{ai} = \mathbf{h}_{a}\mathbf{h}_i^\top
\end{equation}

For the LightGCN model, the model parameters are optimized to minimize the following ranking-based loss $\mathcal{L}_R$:
\begin{equation}\label{eq:ranking_loss}
    \mathcal{L}_R = -\sum_{a\in\mathcal{U}}\sum_{i\in\mathbf{R}^+_a, j\in\mathbf{R}^-_a}log(\sigma(\hat{r}_{ai} - \hat{r}_{aj})),
\end{equation}
where $\sigma(\cdot)$ is the sigmoid function, $\sigma(x)=1/(1+e^{-x})$, $\mathbf{R}^+_a$ denotes the observed items which have interactions with user $a$, and $\mathbf{R}^-_a$ denotes the items which are not connected with the user $a$. 

\subsection{Data Augmentation}
\label{sec:data_augmentation}
According to the setting of the model SGL~\cite{sigir2021SGL}, the interaction data should be disturbed first to generate the augmented user-item bipartite graphs $\mathcal{G}'$ and $\mathcal{G}''$.
In the original paper of SGL, the authors proposed three kinds of data augmentation strategies, \textit{Node Dropout}, \textit{Edge Dropout}, and \textit{Random Walk}. The first two data augmentation strategies randomly drop the nodes and edges of the input user-item bipartite graph by setting the drop ratio $\rho$. The corrupted graphs are also called augmented graphs. And the augmented graphs are fixed among all information propagation layers of the GNN-based module. The \textit{Random Walk} adopts the same strategy as the \textit{Edge Dropout} to augment the input graph, while at different information propagation layers, the augmented graphs are different. More details can refer to the original paper of SGL~\cite{sigir2021SGL}. 
With the augmented graphs, the augmented users' and items' representations $\mathbf{H}'$, $\mathbf{H}''$ can be learned from these augmented data. Finally, an InfoNCE loss is used to push other nodes' representations away from the anchor nodes $i$'s two view representations and pull the two view representations of the same anchor node $i$'s close. The data augmentation strategies which are used in the SGL~\cite{sigir2021SGL} is also used in our proposed model. 

\section{Neighborhood-Enhanced Supervised Contrastive Learning}
\begin{figure*}
\centering
\includegraphics[width=0.9\textwidth]{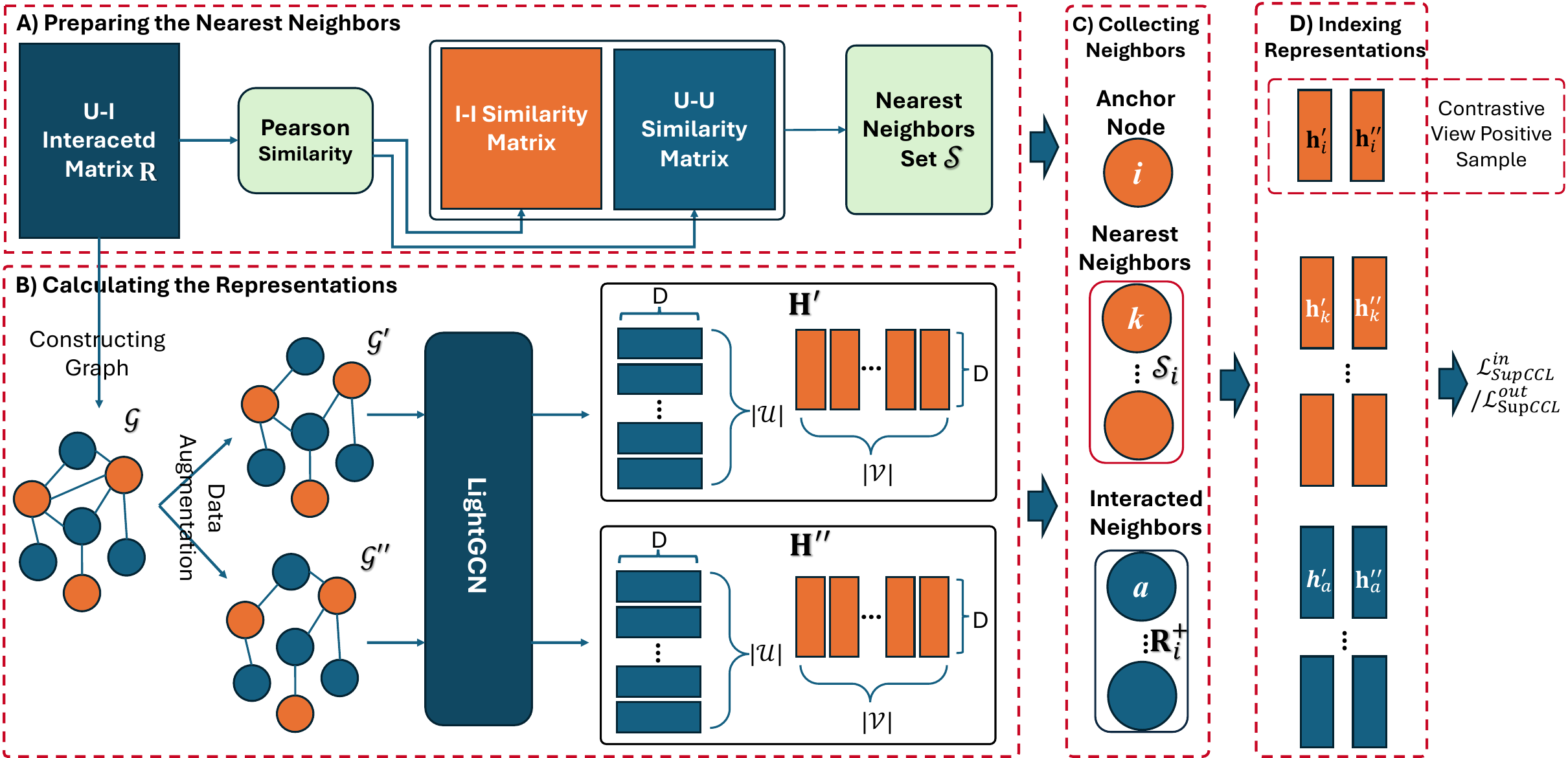}
\caption{The overall framework for utilizing our proposed \fullname. There are four parts, A) It is used to calculate the user-user similarity matrix and item-item similarity matrix based on the user-item interacted matrix $\mathbf{R}$. B) It denotes how to get the two representation matrix $\mathbf{H}'\in\mathbb{R}^{(|\mathcal{U}|+|\mathcal{V}|)\times D}$ and augmented representations $\mathbf{H}''\in\mathbb{R}^{(|\mathcal{U}|+|\mathcal{V}|)\times D}$ of all users and item. The $\mathcal{G}'$ and $\mathcal{G}''$ denote the two augmented graphs, respectively. C) For any anchor node(item $i$), it is necessary to collect its nearest neighbors $\mathcal{S}_i$ based on the item-item similarity matrix and its interacted neighbors based on the user-item interacted matrix. D) Before calculating the supervised collaborative contrastive loss functions $\mathcal{L}^{in}_{NESCL}$ or $\mathcal{L}^{out}_{NESCL}$, we should also index the representations of all users and items from the representation matrix. As the nearest neighbors and interacted neighbors are very clear in this figure, we highlight the contrastive view positive sample of the anchor node in this figure.}
\label{fig:model}
\end{figure*}

This paper aims to modify the traditional contrastive learning technique to incorporate different kinds of positive samples in recommendation tasks. We argue that when constructing the contrastive loss for the anchor node $i$, not only its' representations of two views should be treated as its positive samples, but also the representations of its collaborative neighbors. 
The challenge we address is how to model multiple positive samples of the anchor node.
Inspired by the SupCon~\cite{neurips2021SupCon}, which also focuses on designing the supervised contrastive loss function to model the positive samples. 
We proposed two unique supervised contrastive loss functions \fullname, with ``in''-version and ``out''-version. The two loss functions can refer to Equation~\eqref{eq:NESCL_multiple_positives_in} and Equation~\eqref{eq:NESCL_multiple_positives_out}, respectively. 
The overall framework about how our proposed two unique loss functions work can refer to Figure~\ref{fig:model}. 
We treat the LightGCN~\cite{sigir2020LightGCN} as the backbone model and adopt the same data augmentation strategy as SGL~\cite{sigir2021SGL}. 

The following section will first introduce the preparation for calculating the \shortname. Then, we will 
introduce the forward calculation process.
Last, we will introduce the details of the designed \shortname, and analyze how it dynamically weighs the importance of different kinds of positive samples from the theoretical perspective. 
Finally, we will discuss the complexity of our proposed model and the related model SGL.

\subsection{Preparation for Calculating \shortname}
This section will introduce how to find the anchor node $i$'s nearest neighbors. 
There are two kinds of memory-based methods used in our work, user-based~\cite{ec2000user-based} and item-based~\cite{tois2004ItemKNN}. In this section, we will take the item-based method ItemKNN as an example, introducing how to generate recommendations based on memory-based methods. The recommendation generation procedure can be divided into two sub-procedures. First is calculating the similarity $sim(i, j)$ between any two items $i$ and $j$: 
\begin{equation}
\label{eq:similar_neighbors}
    sim(i, j) = \frac{|\mathbf{R}^+_i\cap\mathbf{R}^+_j|}{\sqrt{|\mathbf{R}^+_i||\mathbf{R}^+_j|}},
\end{equation}
where $|\mathbf{R}^+_i\cap\mathbf{R}^+_j|$ denote how many common interacted users of the items $i$ and $j$. $|\mathbf{R}^+_i|$ and $|\mathbf{R}^+_j|$ denote the degrees of items $i$ and $j$, respectively. 
As the item set $\mathcal{V}$ is very large, to reduce the following time-consuming in generating recommendations, for each item $i$, we treat the top-K items which have the largest $sim(i, j)$ values as $i$'s nearest neighbors. And we use $\mathcal{S}_i$ to denote node $i$'s nearest neighbors set.

\subsection{Model Forward Process}
In the model forward process, we will introduce how to achieve the representations of the anchor node and its positive samples. Then, these achieved representations would be used to calculate the supervised collaborative contrastive loss functions $\mathcal{L}^{in}_{NESCL}$ and $\mathcal{L}^{out}_{NESCL}$. Given the input graph $\mathcal{G}$, it would be augmented twice to get two augmented graphs $\mathcal{G}'$ and $\mathcal{G}''$, with one kind of the data augmentation strategies \textit{Node Dropout}, \textit{Edge Dropout}, and \textit{Random Walk} with drop ratio $\rho$. 
Then, based on the same backbone model LightGCN, we can get two representation matrices, $\mathbf{H}'$ and $\mathbf{H}''$. 
Then, for any anchor node $i$, we index the representations of its nearest neighbors $\mathcal{S}_i$ and interacted neighbors $\mathbf{R}^+_i$. Following, we will introduce the designed loss functions $\mathcal{L}^{in}_{NESCL}$ and $\mathcal{L}^{out}_{NESCL}$ based on the indexed representations. 

\subsection{Details of \shortname}
\label{sec:overall_framwork} 
The main idea of our proposed loss function is, by optimizing the supervised contrastive loss function, the learned anchor's representation should be not only apart from other negative nodes but also close to its collaborative neighbors, i.e., nearest neighbors and interacted neighbors. 
For any anchor node $i$, given its two views of representations $\mathbf{h}'_i$ and $\mathbf{h}''_i$, the representations of its nearest neighbors $\mathbf{h}'_k, k\in\mathcal{S}_i$, and the representations of its interacted neighbors $\mathbf{h}'_a, a\in\mathbf{R}^+_i$, we can get following two kinds of supervised loss functions $\mathcal{L}^{in}_{NESCL}$ and $\mathcal{L}^{out}_{NESCL}$. They are designed to optimize the backbone recommendation model and will work independently.  Our motivation for designing these two loss functions is to investigate the impact of different types of polynomial fusion methods on model optimization under the InfoNCE-based loss function.

The equations of them are:
\begin{align*}
    \mathcal{L}^{in}_{NESCL} = &
    -\sum_{i\in\mathcal{N}}log\frac{exp(\mathbf{h}'_i( \mathbf{h}''_i)^\top/\tau)}{\sum _{j\in\mathcal{N}} exp(\mathbf{h}'_i(\mathbf{h}''_j)^\top/\tau)}
    \\
    &-\sum_{i\in\mathcal{N}}log\sum_{k\in\mathcal{S}_i}\frac{sim(i,k)exp(\mathbf{h}'_k( \mathbf{h}''_i)^\top/\tau)}{\sum _{j\in\mathcal{N}} exp(\mathbf{h}'_k(\mathbf{h}''_j)^\top/\tau)}
    \\
    & -\sum_{i\in\mathcal{N}}log\sum_{a\in\mathbf{R}^+_i}\frac{exp(\mathbf{h}'_a( \mathbf{h}''_i)^\top/\tau)}{\sum _{j\in\mathcal{N}} exp(\mathbf{h}'_a(\mathbf{h}''_j)^\top/\tau)}, \numberthis \label{eq:NESCL_multiple_positives_in}
\end{align*}

and 

\begin{align*}
    \mathcal{L}^{out}_{NESCL} =& 
    -\sum_{i\in\mathcal{N}}log(
    \frac{exp(\mathbf{h}'_i( \mathbf{h}''_i)^\top/\tau)}{\sum _{j\in\mathcal{N}} exp(\mathbf{h}'_i(\mathbf{h}''_j)^\top/\tau)}
    \\
    &+\sum_{k\in\mathcal{S}_i}\frac{sim(i,k)exp(\mathbf{h}'_k( \mathbf{h}''_i)^\top/\tau)}{\sum _{j\in\mathcal{N}} exp(\mathbf{h}'_k(\mathbf{h}''_j)^\top/\tau)}
    \\
    &+\sum_{a\in\mathbf{R}^+_i}\frac{exp(\mathbf{h}'_a( \mathbf{h}''_i)^\top/\tau)}{\sum _{j\in\mathcal{N}} exp(\mathbf{h}'_a(\mathbf{h}''_j)^\top/\tau)}), \numberthis \label{eq:NESCL_multiple_positives_out}
\end{align*}
where the notation $sim(a, i)$ denotes the similarity between the node $a$ and $i$, it is provided by the memory-based methods. And the number of the $\mathcal{S}_i$ is predefined with $K$. We will give more experimental results of K and the influence of neighbors in the experimental part.

Though these two kinds of loss functions seem similar, they play different roles in weighing the importance of different kinds of positive samples. In the following section, we will analyze the difference and how they weigh the difference of different kinds of positive samples from the gradient perspective.

\subsection{Analysis of Our Proposed Loss Functions from the Gradient Perspective}
\label{sec:analysis_gradient}

\subsubsection{Analysis of the ``in''-version Loss Function $\mathcal{L}^{in}_{NESCL}$}
\label{sec:analysis_loss_in_NESCL}
To better study the proposed contrastive loss function, first, we use the $\mathcal{L}^{in}_{NESCL}$ to the following equation:
\begin{align*}
    \mathcal{L}^{in}_{NESCL} = &
    -log\frac{exp(\mathbf{h}'_i( \mathbf{h}''_i)^\top/\tau)}{\sum _{j\in\mathcal{N}} exp(\mathbf{h}'_i(\mathbf{h}''_j)^\top/\tau)}
    \\
    &-log\frac{exp(\mathbf{h}'_k( \mathbf{h}''_i)^\top/\tau)}{\sum _{j\in\mathcal{N}} exp(\mathbf{h}'_k(\mathbf{h}''_j)^\top/\tau)}
    \\
    & -log\frac{exp(\mathbf{h}'_a( \mathbf{h}''_i)^\top/\tau)}{\sum _{j\in\mathcal{N}} exp(\mathbf{h}'_a(\mathbf{h}''_j)^\top/\tau)}, \numberthis \label{eq:NESCL_multiple_positives_inx}
\end{align*}

Then, we calculate the gradient from $\mathcal{L}^{in}_{NESCL}$ to the anchor node $i$'s representation $\mathbf{h}''_i$. We can get the following equation:
\begin{equation}
\frac{\partial \mathcal{L}_{NESCL}^{in}}{\partial \mathbf{h}''_i}=\underbrace{\lambda_i^{in} \mathbf{h}_i^{\prime}}_{S G L}+\underbrace{\lambda_k^{i n} \mathbf{h}_k^{\prime}+\lambda_a^{i n} \mathbf{h}_a^{\prime}}_{\text {Neighborhood-Enhanced }}.
\end{equation}

According to the analysis of SGL~\cite{sigir2021SGL}, we highlight the difference between our proposed loss function and the loss function in SGL.
From the above formula, we can find, with the help of the Neighborhood-Enhanced term, the anchor node's embedding $\mathbf{h}''_i$ is decided by the positive samples $\mathbf{h}'_i$, $\mathbf{h}'_k$, and $\mathbf{h}'_a$ simultaneously. It is one of the reasons why our proposed loss function can better guide the optimization of the backbone model. 

Second, the influence capacity $\lambda^{in}_i$, $\lambda^{in}_k$, and $\lambda^{in}_a$ of different positive samples are decided by the anchor node's representation $\mathbf{h}''_i$ and many negative samples $(\mathbf{h}''_j, j\in\mathcal{N},j\neq i)$. It makes the computation of the influence capacity more accurate. 
For example, the value $\lambda^{in}_k$ can be calculated with:
\begin{align*}
    \lambda^{in}_{k} = \frac{1}{\tau}(\frac{-\sum_{j\in\mathcal{N},j\neq i}exp(\mathbf{h}^{'}_{k}(\mathbf{h}^{''}_j)^{\top}/\tau)}{exp(\mathbf{h}^{'}_{k}(\mathbf{h}^{''}_i)^{\top}/\tau)+\sum_{j\in\mathcal{N},j\neq i}exp(\mathbf{h}^{'}_{k}(\mathbf{h}^{''}_j)^{\top}/\tau)}).\numberthis
\end{align*}

\subsubsection{Analysis of the ``out''-version Loss Function $\mathcal{L}^{out}_{NESCL}$}
Similar to the analysis in the last subsection, we adopt the same method to analyze the loss function $\mathcal{L}^{out}_{NESCL}$. By calculating the gradient of $\mathcal{L}^{out}_{NESCL}$ to the node $i$'s auxiliary view $\mathbf{h}_i''$, we can get:
\begin{align*}
    \frac{\partial\mathcal{L}^{out}_{NESCL}}{\partial\mathbf{h}''_i}=\lambda^{out}_i\mathbf{h}'_i + \lambda^{out}_{k}\mathbf{h}'_{k}+\lambda^{out}_{a}\mathbf{h}'_{a}. , \numberthis \label{eq:xxx}
\end{align*}

According to the above formula, we can get the same conclusion as the $\mathcal{L}^{in}_{NESCL}$. And the difference between these two kinds of loss functions is the calculated influence capacity of different positive samples. Compared with the ``in''-version loss function, the computation of the ``out''-version would be more complex. 
As the formula of the $\lambda^{out}_i$, $\lambda^{out}_{k}$, and $\lambda^{out}_{a}$ are pretty complex, we would not expand them here. Please refer to Appendix sections A.1 and A.2 for more details. 
We take the $\lambda^{out}_{k}$ as an example.
By dividing the $\lambda^{in}_k$ by $\lambda^{out}_{k}$, we can get the following:
\begin{align*}
    \frac{\lambda^{in}_k}{\lambda^{out}_k} =
    1
    &+\frac{1+\frac{\sum_{j\in\mathcal{N},j\neq i}exp(\mathbf{h}^{'}_k(\mathbf{h}^{''}_j)^{\top}/\tau)}{exp(\mathbf{h}^{'}_k(\mathbf{h}^{''}_i)^{\top}/\tau)}}{1+\frac{\sum_{j\in\mathcal{N},j\neq i}exp(\mathbf{h}^{'}_{i}(\mathbf{h}^{''}_j)^{\top}/\tau)}{exp(\mathbf{h}^{'}_{i}(\mathbf{h}^{''}_i)^{\top}/\tau)}}
    \\
    &+\frac{1+\frac{\sum_{j\in\mathcal{N},j\neq i}exp(\mathbf{h}^{'}_k(\mathbf{h}^{''}_j)^{\top}/\tau)}{exp(\mathbf{h}^{'}_k(\mathbf{h}^{''}_i)^{\top}/\tau)}}{1+\frac{\sum_{j\in\mathcal{N},j\neq i}exp(\mathbf{h}^{'}_{a}(\mathbf{h}^{''}_j)^{\top}/\tau)}{exp(\mathbf{h}^{'}_{a}(\mathbf{h}^{''}_i)^{\top}/\tau)}},\numberthis
\end{align*}

From the formula, we have two observations. First is, the value of $\lambda^{out}_i$ should be smaller than $\lambda^{in}_i$. 
Second, compared with $\lambda^{in}_i$, the value of $\lambda^{out}_i$ is not only affected by the distance between its corresponding positive sample $\mathbf{h}_i'$, but also the other positive samples $\mathbf{h}'_{k}$, and $\mathbf{h}'_{a}$. 
We would evaluate the performance of these two kinds of loss functions in the experimental section.

\subsection{Overall Loss Functions of Our Proposed Model}
In this section, we introduce the overall loss function of our proposed model. Although the supervised collaborative contrastive loss we proposed can utilize the information of different kinds of positive samples in the training stage, when conducting experiments, we found that the loss function $\mathcal{L}_{R}$ in Equation ~\eqref{eq:ranking_loss} is also helpful. We suppose that the two kinds of loss functions can provide different kinds of capacity to pull the anchor node and the positive samples close in the representation space. 
Thus, the overall loss functions of our proposed model are:
\begin{align*}
    \mathcal{L}^{in}_{\mathcal{O}} &= \mathcal{L}_R + \alpha\mathcal{L}^{in}_{NESCL},
    \\
    \mathcal{L}^{out}_{\mathcal{O}} &= \mathcal{L}_R + \alpha \mathcal{L}^{out}_{NESCL},
    \numberthis \label{eq:NESCL_overall_loss}
\end{align*}
where $\alpha$ is a hyper-parameter to balance the importance of the two kinds of loss functions. Larger $\alpha$ means the corresponding loss plays a more important role in the training stage.

\subsection{Time Complexity}
In this subsection, we mainly analyze the time complexity when utilizing our proposed loss functions.
The overall framework contains the following modules, data preparation, data augmentation, graph convolution, ranking-based loss calculation, and supervised collaborative contrastive loss calculation. 
As the data augmentation, graph convolution, and ranking-based loss calculation modules are the same as the SGL model, we wouldn't discuss them here. 

The time-consuming procedure for the nearest neighbors finding operation is calculating the user-user similarity and item-item similarity matrices. Its time complexity is $O(|\mathcal{U}||\mathcal{U}|)+O(|\mathcal{V}||\mathcal{V}|)$. Though the time complexity is high, we only calculate the similarity matrices once. However, in practical applications, we may use different algorithms to discover nearest neighbors with varying time complexities. This step should be considered as a part of the data preparation stage, and thus, we will not discuss its impact on the training complexity of our model.

The time complexity of the $\mathcal{L}^{in}_{NESCL}$ and $\mathcal{L}^{out}_{NESCL}$ should be the same. And we take the $\mathcal{L}^{in}_{NESCL}$ as an example. 
For the first term in Equation~\eqref{eq:NESCL_multiple_positives_in}, we can easily find that the complexity of the numerator and denominator should be $O(|\mathcal{N}|D)$ and $O(|\mathcal{N}||\mathcal{N}|D)$, respectively. As we only treat other nodes in the batch as the negative samples, the time complexity of the 
Thus, the time complexity of the denominator can be corrected as $O(|\mathcal{N}|BD)$), where $B$ is the batch size.  
Similarly, the time complexity of the second term should be $(O(K|\mathcal{N}|D)$ + $O(K|\mathcal{N}|BD))$, where $K$ is the number of nearest neighbors for each anchor node $i$. However, in practice, we found that randomly selecting one nearest neighbor from the neighbor set $\mathcal{S}_i$ may achieve better performance. Thus, the time complexity of the second term can be reduced to $(O(|\mathcal{N}|D)$ + $O(|\mathcal{N}|BD))$. 
For the third term in Equation~\eqref{eq:NESCL_multiple_positives_in}, the time complexity should be $(O(|\mathcal{E}|D)$ + $O(2|\mathcal{E}|BD))$, the number $2$ means all user-item pairs would appear twice in the third term calculation procedure. 
The overall supervised collaborative contrastive loss function is 
$O(|\mathcal{N}|D(2+2B))s$ +
$O(|\mathcal{E}|D(1+2B))s$.

\begin{table}[]
\caption{Time Complexity of the Overall Framework for Our Proposed Loss Function and SGL. ($D$ denotes the vector dimension size, $\rho$ is the data drop ratio in the data augmentation procedure, $s$ is the training epoch number, $B$ is the batch size, and $L$ is the GNN propagation layer number.)}
\scriptsize
\begin{tabular}{|l|lc|}
\hline
Component                                                                            & \multicolumn{1}{c|}{SGL}                                               & \shortname                                                                                                                                                                                                                          \\ \hline
Adjacency Matrix                                                                     & \multicolumn{2}{c|}{$O(4 \rho|\mathcal{E}| s+2|\mathcal{E}|)$}                                                                                \\ \hline
Graph Convolution                                                                    & \multicolumn{2}{c|}{$O(2(1+2 \rho)|\mathcal{E}| LD \frac{|\mathcal{E}|}{B}) s $}                                                                      \\ \hline
BPR Loss                                                                             & \multicolumn{2}{c|}{$O(2|\mathcal{E}| D s)$}                                                                                                                                                                                              \\ \hline
Self-supervised Loss                                                                 & \multicolumn{1}{c|}{$O(|\mathcal{N}| D(1+ B) s)$} & -                                                                                                                                                                                                                               \\ \hline
\begin{tabular}[c]{@{}l@{}}Supervised Collaborative \\ Contrastive Loss\end{tabular} & \multicolumn{1}{c|}{-}                                                 & \begin{tabular}[c]{@{}l@{}}$O(|\mathcal{N}|D(2+2B))s$ +\\ $O(|\mathcal{E}|D(1+2B))s$\end{tabular} \\ \hline
\end{tabular}
\end{table}

\section{Experiment}
In the experiment section, we aim to answer the following two questions. 
\begin{itemize}
    \item \textbf{RQ1}: Can our proposed supervised loss functions help the backbone model perform better?
    \item \textbf{RQ2}: How about the performance of the backbone model under variants of our proposed loss functions?
\end{itemize}
Following, we first introduce the experiment settings, and then we will answer the above questions individually. As there are some hyper-parameters not important for verifying the effectiveness of our proposed loss function, we would report the results that are related to them in Appendix Section B. 

\subsection{Datasets and Metrics}
We conduct experiments based on three public real-world datasets, Yelp2018, Gowalla, and Amazon-Book, which are provided by the authors of LightGCN~\cite{sigir2020LightGCN} in the link\footnote{\url{https://github.com/kuandeng/LightGCN}}. 
These datasets contain the user-item interacted records.
The statistics of these datasets can refer to Table~\ref{tab:stat}. To keep the results the same as the authors reported in these works~\cite{sigir2020LightGCN, sigir2021SGL, cikm2021SimpleX, cikm2021UltraGCN}, we also utilize the original format of the provided datasets without any modification. 

\begin{table}[htb]
\centering
\caption{Statistics of the Three Real-World Datasets.}\label{tab:stat}
\small{
\begin{tabular}{c|l|l|l|l}
\hline
Datasets                 & Users   & Items   & Ratings   & Density \\ \hline
Amazon-Book & 55,188 & 9,912 & 1,445,622 & 0.062\% \\ 
\hline
Gowalla &  29,858 & 40,981 & 1,027,370 & 0.084\% \\ \hline
Yelp2018 &  31,688 & 38,048 & 1,561,406 & 0.130\% \\ \hline
\end{tabular}
}
\end{table}

\textbf{Metrics} In this study, we use the metrics Recall@K and NDCG@K to evaluate the performance of all models~\cite{sigir2021SGL}. $K$ denotes only top-K recommended items for each user are assessed. Recall@K measures how many of a user's interacted items appear in the recommendation list. NDCG@K measures whether the user's interacted products rank first in the recommendation list. Larger Recall@K and NDCG@K mean better performance. And $K$ is fixed as 20. We implement the backbone model and our proposed loss functions based on the RecBole recommendation library~\cite{cikm2021RecBole}. 

\textbf{Baselines.}
We focus on the candidate-matching task, and the data we used is made of the user-item interacted records, thus we select several classical latent factor-based collaborative filtering models as the baseline models. As the backbone for our proposed loss function is the GNN-based models, thus we also treat the SOTA GNN-based collaborative filtering models as the baseline models. Last, the contrastive loss-based model SGL is also be treated as the baseline model, as it is the SOTA model which utilizes the contrasitve loss function. 
Although the SupCon is designed for the computer vision task, it is similar to the loss function we proposed. We also modify the origin SupCon to make it can be used in the recommendation task. 
Besides, as we also utilize memory-based methods to find the nearest neighbors, we also test the performance of some classical memory-based methods. 
And the details of the baseline models are as follows. 

\textbf{Group 1: Latent Factor based CF Models.}
\textbf{BPR} is a competitive classical recommendation model~\cite{uai2012BPR}. It is proposed to model the relationship between any positive user-item pair and negative user-item pair.
\textbf{SimpleX}~\cite{cikm2021SimpleX} is proposed to advance the interaction encoder module, loss function of current candidate matching models, such as BPR~\cite{uai2012BPR}, LightGCN~\cite{sigir2020LightGCN}, and so on.

\textbf{Group 2: Graph Neural Network based CF Models.}
\textbf{LightGCN}~\cite{sigir2020LightGCN} is a simplified version of the graph convolution network based recommender model NGCF~\cite{sigir2019NGCF}. It removes the heavy trainable transform variables and the non-linear activation function of NGCF.
\textbf{UltraGCN} models not only the user-item relationship but also the item-item relationship~\cite{cikm2021UltraGCN}. And it achieves competitive performance. 

\textbf{Group 3: Contrastive Learning based CF Models.}
\textbf{SGL} utilizes the contrasitve learning technique to eliminate the noise which is brought by the LightGCN model~\cite{sigir2021SGL}. As the \textit{Edge Dropout} makes SGL performs best among all datasets, we also only introduce the result of \textbf{SGL(ED)} model. 
\textbf{SupCon}: It is a kind of supervised contrastive loss function. In the original paper of SupCon, the researchers proposed two kinds of loss functions. Though they are designed for the computer vision task, we directly follow the design in its original manuscript to design the loss functions for the recommendation task.
We use SupCon(in) and SupCon(out) to denote training the model with minimizing the $\mathcal{L}^{in}_{SupCon}$ and $\mathcal{L}^{out}_{SupCon}$, respectively. 

\textbf{Group 4: Memory-based CF Models.}
\textbf{User-based} is a classical memory-based CF model~\cite{ec2000user-based} to find users' collaborative neighbors according to the users' historical interaction records. 
\textbf{ItemKNN} is similar to the user-based CF model~\cite{tois2004ItemKNN}. It is also a simple but effective memory-based CF model. It is used to find the items' nearest neighbors. In this paper, we also use the User-based~\cite{ec2000user-based} and ItemKNN~\cite{tois2004ItemKNN} to find the anchor node's nearest neighbors. 

\subsection{Parameter Settings}
In this section, we mainly introduce the setting of the parameters in our work. 
The regularization value $\alpha$ of Equation~\eqref{eq:NESCL_overall_loss} are set to 0.3, 0.1, and 0.3 for Yelp2018, Gowalla, and Amazon-Book respectively. The data augmentation ratio $\rho$ is set to 0.3 for all datasets, which is introduced in section \ref{sec:data_augmentation}. 
And we adopt the data augmentation with \textit{Node Dropout}, \textit{Node Dropout}, and \textit{Edge Dropout} for Yelp2018, Gowalla, and Amazon-Book, respectively. 
Please note that the number of negative samples of all baseline models is set to 1 but for the SimpleX and UltraGCN models. According to the official implementation of UltraGCN\footnote{\url{https://github.com/xue-pai/UltraGCN}}, the number of negative samples is set to 800, 1500, and 500 for Yelp2018, Gowalla, and Amazon-Book datasets, respectively. And for the SimpleX\footnote{\url{https://github.com/xue-pai/TwoTowers/blob/master/benchmarks/Yelp18/}}, the number of negative samples for Yelp2018 is set to 1000, while the configure files for Gowalla and Amazon-Book are not provided. For more details about implementing the overall framework can refer to the following link\footnote{\url{https://gitee.com/peijie_hfut/nescl}}.

Inspired by GACL~\cite{sigir2022GACL}, the model removing the initial embedding of LightGCN performs better on Yelp2018 and Gowalla. While on Amazon-Book, removing initial embedding performs worse. Thus for the Amazon-Book, we would keep the initial embedding of LightGCN, and remove it for Yelp2018 and Gowalla.

\begin{table*}[htb]
\centering
\caption{Overall Performance Among All Models On Three Real-World Datasets.}
\label{tab:overall}
\begin{tabular}{|c|rr|rr|rr|}
\hline
Datasets & \multicolumn{2}{c|}{Yelp2018}                                 & \multicolumn{2}{c|}{Gowalla}                                  & \multicolumn{2}{c|}{Amazon-Book}                              \\ \hline
Model    & \multicolumn{1}{l|}{Recall@20} & \multicolumn{1}{l|}{NDCG@20} & \multicolumn{1}{l|}{Recall@20} & \multicolumn{1}{l|}{NDCG@20} & \multicolumn{1}{l|}{Recall@20} & \multicolumn{1}{l|}{NDCG@20} \\ \hline
User-based  & \multicolumn{1}{r|}{0.0463}    & 0.0397                       & \multicolumn{1}{r|}{0.1035}    & 0.0815                       & \multicolumn{1}{r|}{0.0329}    & 0.0295                       \\ \hline
ItemKNN  & \multicolumn{1}{r|}{0.0639}    & 0.0531                       & \multicolumn{1}{r|}{0.1570}    & 0.1214                       & \multicolumn{1}{r|}{\textbf{0.0736}}    & \textbf{0.0606}                       \\ \hline \hline
BPR      & \multicolumn{1}{r|}{0.0576}    & 0.0468                       & \multicolumn{1}{r|}{0.1627}    & 0.1378                       & \multicolumn{1}{r|}{0.0338}    & 0.0261                       \\ \hline
LightGCN & \multicolumn{1}{r|}{0.0649}    & 0.0530                       & \multicolumn{1}{r|}{0.1830}    & 0.1550                       & \multicolumn{1}{r|}{0.0411}    & 0.0315                       \\ \hline
UltraGCN & \multicolumn{1}{r|}{0.0683}    & 0.0561                       & \multicolumn{1}{r|}{0.1862}    & 0.1580                       & \multicolumn{1}{r|}{0.0681}    & 0.0556                       \\ \hline
SimpleX  & \multicolumn{1}{r|}{0.0701}    & 0.0575                       & \multicolumn{1}{r|}{0.1872}    & 0.1557                       & \multicolumn{1}{r|}{0.0583}    & 0.0468                       \\ \hline \hline
SGL   & \multicolumn{1}{r|}{0.0675}    & 0.0555                       & \multicolumn{1}{r|}{0.1787}    & 0.1510                       & \multicolumn{1}{r|}{0.0478}    & 0.0379                       \\ \hline
SupCon(in)   & \multicolumn{1}{r|}{0.0727}    & 0.0599                       & \multicolumn{1}{r|}{0.1900}    & 0.1607                       & \multicolumn{1}{r|}{0.0616}    & 0.0505                       \\ \hline
SupCon(out)   & \multicolumn{1}{r|}{0.0739}    & 0.0609                       & \multicolumn{1}{r|}{0.1897}    & 0.1605                       & \multicolumn{1}{r|}{0.0339}    & 0.0288                       \\ \hline
\hline
\shortname(in)  & \multicolumn{1}{r|}{0.0732}    & 0.0602                       & \multicolumn{1}{r|}{0.1913}    & \textbf{0.1617}                       & \multicolumn{1}{r|}{0.0624}    & 0.0513                       \\ \hline
\shortname(out)  & \multicolumn{1}{r|}{\textbf{0.0743}}    & \textbf{0.0611}                       & \multicolumn{1}{r|}{\textbf{0.1917}}    & \textbf{0.1617}                       & \multicolumn{1}{r|}{0.0483}    & 0.0379                       \\ \hline
\end{tabular}
\end{table*}
\subsection{Overall Analysis of Our Proposed Loss Functions(RQ1)}
In this section, we would like to answer the research question RQ1, i.e., how about the performance of the backbone model which is trained upon our proposed supervised loss function? 

\subsubsection{Overall Performance of All Baseline Models}
\label{sec:overall_performance}
The experimental results of all baseline models are copied from this web page\footnote{\url{https://openbenchmark.github.io/candidate-matching/}}, which are built by the authors of SimpleX~\cite{cikm2021SimpleX} and UltraGCN~\cite{cikm2021UltraGCN}. We have double-checked part of the results, and they are consistent with the original papers. 
We report the performance of the backbone model LightGCN under two kinds of objective loss functions. The \shortname(in) denotes training the backbone model by minimizing the $\mathcal{L}^{in}_{\mathcal{O}}$ in Equation~\eqref{eq:NESCL_overall_loss}. And \shortname(out) denotes training the backbone model by minimizing the $\mathcal{L}^{out}_{\mathcal{O}}$ in Equation~\eqref{eq:NESCL_overall_loss}.
Please note that, when training the model for the Amazon-Book dataset, the $\mathcal{L}_R$ should be removed in Equation~\eqref {eq:NESCL_overall_loss}. We have analyzed the reason in section~\ref{ref:combination}. 
The overall performance of all models can refer to Table~\ref{tab:overall}. From the experimental results, we have the following three observations:

1. From the experimental results, we can find the backbone which is trained based on our proposed loss functions outperforms all baseline models on both Yelp2018 and Gowalla datasets, especially the contrastive learning based models, such as SGL, SupCon(in), and SupCon(out). On Gowalla dataset, the performance of SGL is inferior to LightGCN on Gowalla. It may be because the contrastive loss may destroy the power of the ranking loss to pull interacted neighbors close in the representation space. Compared with the two versions of SupCon, our proposed loss function outperform them, which can also verify that our proposed loss function can better incorporate different positive samples. On different datasets, the ``in''-version and ``out''-version of our proposed loss functions perform inconsistently on all datasets. We also select the best version of the loss function for each dataset in the following experiments. 

2. SimpleX and UltraGCN outperform other baseline models on all datasets. The SimpleX model adopts a novel contrastive loss function to model the relationship between positive samples and negative samples in the training stage. Based on the new contrastive loss function, increasing the number of negative samples can improve the performance of the SimpleX a lot. As for other latent factor-based models, such as BPR, LightGCN, SGL, they all used the ranking loss function in Equation~\eqref{eq:ranking_loss}, and the number of negative samples is set to 1. The reason why the UltraGCN outperforms other baseline models may be because it incorporates nearest neighbors. 

3. From Table~\ref{tab:overall}, it is surprising that the memory-based models perform much better than the latent factor model BPR on all datasets. Especially on the Amazon-Book dataset, the classical memory-based model SLIM outperforms all other models. As our proposed loss function can be used to incorporate the nearest neighbors and latent factor-based model meantime, the backbone model also outperforms the memory-based models and several latent factor models a lot except the Amazon-Book dataset. 
We think the possible reason may be the exposure bias in the Amazon-Book dataset. As in the amazon online shopping website, the recommender system prefers the item-based memory method to provide the item recommendation list for the customers. Thus the models that incorporate the nearest neighbors would achieve a nice performance, such as SLIM, UltraGCN, and our work.

\begin{table}[htb]
\scriptsize
\caption{Different Objective Loss Functions~(Taking the $\mathcal{L}^{in}_{NESCL}$(Equation~\eqref{eq:NESCL_multiple_positives_in}) as an Example).}
\label{tab:intro_loss_funcs}
\begin{tabular}{|c|c|}
\hline
Optimization Term & Equation \\ \hline
Ranking Loss & $\mathcal{L}_R$ \\ \hline
Different Views & $
    -\sum_{i\in\mathcal{N}}log\frac{exp(\mathbf{h}'_i( \mathbf{h}''_i)^\top/\tau)}{\sum _{j\in\mathcal{N}} exp(\mathbf{h}'_i(\mathbf{h}''_j)^\top/\tau)}$ \\ \hline
Interacted Neighbors & $-\sum_{i\in\mathcal{N}}log\sum_{a\in\mathbf{R}^+_i}\frac{exp(\mathbf{h}'_a( \mathbf{h}''_i)^\top/\tau)}{\sum _{j\in\mathcal{N}} exp(\mathbf{h}'_a(\mathbf{h}''_j)^\top/\tau)}$  \\ \hline
User Nearest Neighbors & $-\sum_{i\in\mathcal{U}}log\sum_{k\in\mathcal{S}_i}\frac{sim(i,k)exp(\mathbf{h}'_k( \mathbf{h}''_i)^\top/\tau)}{\sum _{j\in\mathcal{N}} exp(\mathbf{h}'_k(\mathbf{h}''_j)^\top/\tau)}$  \\ \hline
Item Nearest Neighbors & $-\sum_{i\in\mathcal{V}}log\sum_{k\in\mathcal{S}_i}\frac{sim(i,k)exp(\mathbf{h}'_k( \mathbf{h}''_i)^\top/\tau)}{\sum _{j\in\mathcal{N}} exp(\mathbf{h}'_k(\mathbf{h}''_j)^\top/\tau)}$  \\ \hline
All & $\mathcal{L}_{NESCL}$+$\mathcal{L}_R$ \\ \hline
\end{tabular}
\end{table}

\begin{table*}[]
\centering
\caption{The Performance of Our Proposed Supervised Contrastive Loss Function Incorporating Different Kinds of Positive Samples.(The following optimization terms are modified based on the Equation~\eqref{eq:NESCL_overall_loss}, which corresponds to the ``All'' term. 
``Ranking Loss'' denotes the $\mathcal{L}_R$ loss, ``Different Views'' denotes only two views of representations are treated as positive samples. ``Interacted Neighbors'' denotes only the anchor nodes' interacted neighbors are treated as positive samples. ``User Nearest Neighbors'' denotes only the users' nearest neighbors are treated as positive samples. ``Item Nearest Neighbors'' denotes only the items' nearest neighbors are treated as positive samples. ``+'' operation denotes optimizing the summed two kinds of loss functions. And the ``-'' operation denotes only the latter loss that is not considered.)}
\label{tab:different_modules}
\begin{tabular}{|c|ll|ll|ll|}
\hline
\multirow{2}{*}{Optimization}          & \multicolumn{2}{c|}{Yelp2018}            & \multicolumn{2}{c|}{Gowalla}             & \multicolumn{2}{c|}{Amazon-Book}         \\ \cline{2-7} 
                                & \multicolumn{1}{l|}{Recall@20} & NDCG@20 & \multicolumn{1}{l|}{Recall@20} & NDCG@20 & \multicolumn{1}{l|}{Recall@20} & NDCG@20 \\ \hline
Ranking Loss          & \multicolumn{1}{l|}{0.0649}    & 0.0530  & \multicolumn{1}{l|}{0.1830}    & 0.1550  & \multicolumn{1}{l|}{0.0411}    & 0.0315  \\ \hline
Different Views             & \multicolumn{1}{l|}{0.0434}    & 0.0362  & \multicolumn{1}{l|}{0.0784}    & 0.0570  & \multicolumn{1}{l|}{0.0063}    & 0.0051  \\ \hline
Ranking + Different Views               & \multicolumn{1}{l|}{0.0675}    & 0.0555  & \multicolumn{1}{l|}{0.1787}    & 0.1510  & \multicolumn{1}{l|}{0.0478}    & 0.0379  \\ \hline
Interacted Neighbors        & \multicolumn{1}{l|}{0.0682}    & 0.0566  & \multicolumn{1}{l|}{0.1774}    & 0.1496  & \multicolumn{1}{l|}{0.0503}    & 0.0395  \\ \hline
User Nearest Neighbors       & \multicolumn{1}{l|}{0.0544}    & 0.0456  & \multicolumn{1}{l|}{0.1065}    & 0.0878  & \multicolumn{1}{l|}{0.0277}    & 0.0217  \\ \hline
Item Nearest Neighbors       & \multicolumn{1}{l|}{0.0529}    & 0.0441  & \multicolumn{1}{l|}{0.1062}    & 0.0799  & \multicolumn{1}{l|}{0.0284}    & 0.0247  \\ \hline
Nearest Neighbors & \multicolumn{1}{l|}{0.0600}    & 0.0502  & \multicolumn{1}{l|}{0.1073}    & 0.0829  & \multicolumn{1}{l|}{0.0398}    & 0.0332  \\ \hline
Interacted Neighbors + Nearest Neighbors  & \multicolumn{1}{l|}{0.0717}    & 0.0594  & \multicolumn{1}{l|}{0.1778}    & 0.1493  & \multicolumn{1}{l|}{0.0609}    & 0.0497  \\ \hline
All - Ranking Loss & \multicolumn{1}{l|}{0.0705}    & 0.0586  & \multicolumn{1}{l|}{0.1741}    & 0.1444  & \multicolumn{1}{l|}{\textbf{0.0624}}    & \textbf{0.0513}  \\ \hline
All  & \multicolumn{1}{l|}{\textbf{0.0743}}    & \textbf{0.0611}  & \multicolumn{1}{l|}{\textbf{0.1917}}    & \textbf{0.1617}  & \multicolumn{1}{l|}{0.0580}    & 0.0473  \\ \hline
\end{tabular}
\end{table*}

\subsection{Analysis of NESCL(RQ2)}
In this section, we would analyze the important parameters in optimizing our proposed loss functions. As our proposed loss function contains both the supervised collaborative contrastive loss function and ranking loss in Equation~\eqref{eq:NESCL_overall_loss}, we incorporate several kinds of positive samples in the supervised contrastive loss function. In this section, we would like to test the influence of different kinds of positive samples and loss functions on model training. We argued that our proposed loss function can better utilize the negative samples which are not hard, especially when the temperature value $\tau$ is small. Thus, we test the performance of our proposed loss functions under different temperature values in the second subsection. Third, we will study how to incorporate the nearest neighbors in the supervised contrastive loss function, especially when the quality of the nearest neighbors is not guaranteed.
Besides, there are some other hyper-parameters that are not important in verifying the key ideas of our proposed loss functions, we only report the results which are related to such hyper-parameters in Appendix section B. 

\subsubsection{Different Combinations of Loss Functions}
\label{ref:combination}
In this section, we will report the results of our proposed loss functions on three real-world datasets when minimizing different combinations of loss functions. The experiment results can refer to Table~\ref{tab:different_modules}, more details about these loss functions can refer to Table~\ref{tab:intro_loss_funcs}. In Table~\ref{tab:different_modules}, each term of the ``Optimization'' column denotes the loss which is minimized. The backbone model achieves the best performance with $\mathcal{L}^{in}_{NESCL}$ on Yelp2018 and Gowalla, and achieves best performance on Amazon-Book with $\mathcal{L}^{out}_{NESCL}$. The ``Ranking + Different Views'' corresponds to the performance of SGL.
We have the following observations:

1.  The performance of the model is very worse when only optimizing the ``Different Views'' loss, i.e., only treating the representations of two views as positive samples. Thus the ranking loss $\mathcal{L}_R$ is necessary. It means that only pushing apart the anchor node with any other nodes in the representation space will lead to worse performance. 

2. On Yelp2018 and Amazon-Book, the performance of the model obtained by optimizing only ``Interacted Neighbors''  even exceeds the model obtained by optimizing the ranking loss $\mathcal{L}_R$. It may be because the contrastive loss with a small temperature can provide larger gradient values. By only optimizing the ranking loss, the backbone model still performs very well on Gowalla dataset, and by adding our proposed supervised contrastive loss to the ranking loss, the backbone model can achieve better performance, which is shown in the ``All'' row. 

3. Our proposed ``Interacted Neighbors'' loss and ``Nearest Neighbors'' loss can address of the limitation of SGL, i.e., a small temperature of the contrastive loss may destroy the ability of the ranking loss to pull any interacted nodes close in the representation space. Such a conclusion can be gotten by comparing the results of ``All - Ranking Loss'', ``Different Views'', and ``User \& Item Nearest Neighbors''.
On the amazon-book dataset, only optimizing the ``Ranking Loss'' is inferior to only optimizing the ``Interacted Neighbors'', and we found removing the ``Ranking Loss'' from the ``All'' loss can make the model perform better. However, we suppose optimizing the ``Ranking Loss'' may hurt the representations learning of the backbone model, which such issue can be alleviated by our proposed ``Interacted Neighbors'' loss. Unfortunately, we don't know why such a phenomenon appears on Amazon-Book dataset.

\begin{figure*}[htb]
\centering
\subfloat[Recall@20 on Yelp2018]{
    \begin{minipage}[b]{0.3\textwidth}
    \includegraphics[width=1\textwidth]{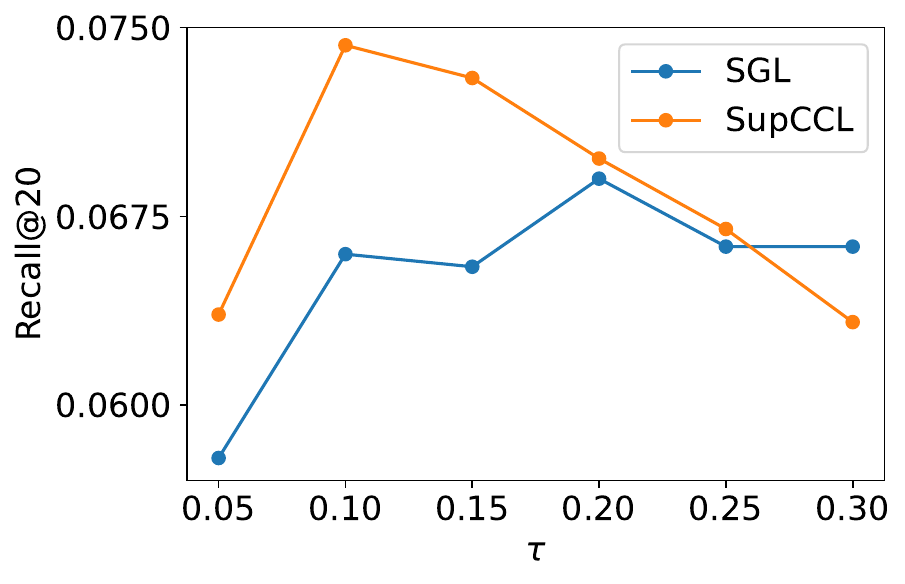}
    \end{minipage}
}
\subfloat[Recall@20 on Gowalla]{
    \begin{minipage}[b]{0.3\textwidth}
    \includegraphics[width=1\textwidth]{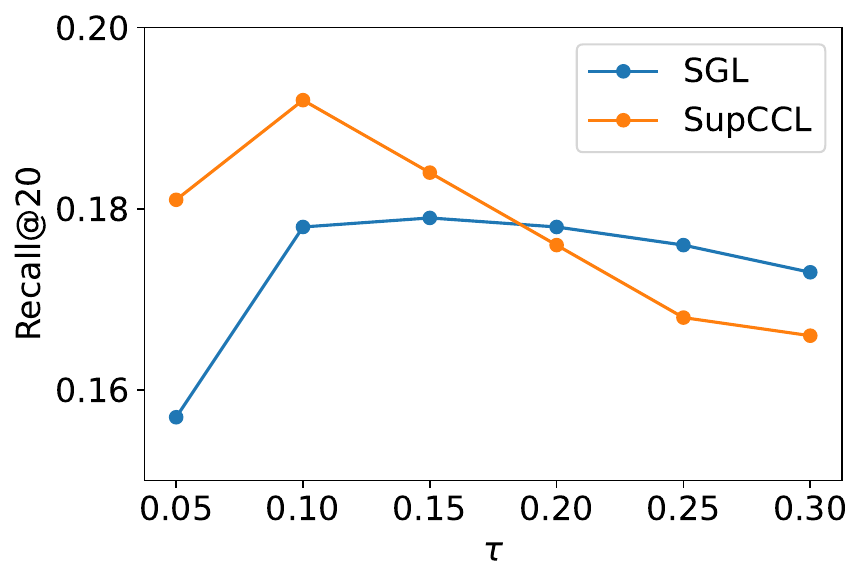}
    \end{minipage}
}
\subfloat[Recall@20 on Amazon-Book]{
    \begin{minipage}[b]{0.3\textwidth}
    \includegraphics[width=1\textwidth]{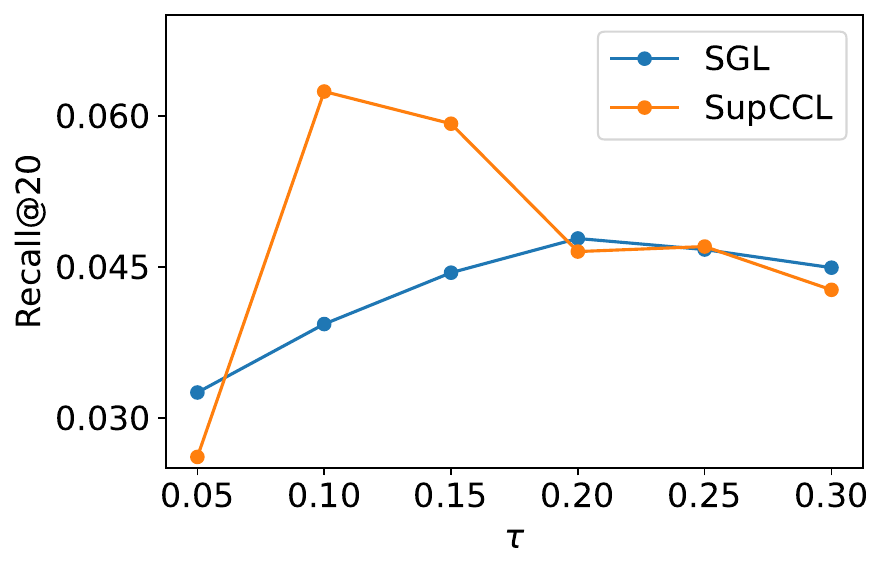}
    \end{minipage}
}

\quad
\subfloat[NDCG@20 on Yelp2018]{
    \begin{minipage}[b]{0.3\textwidth}
    \includegraphics[width=1\textwidth]{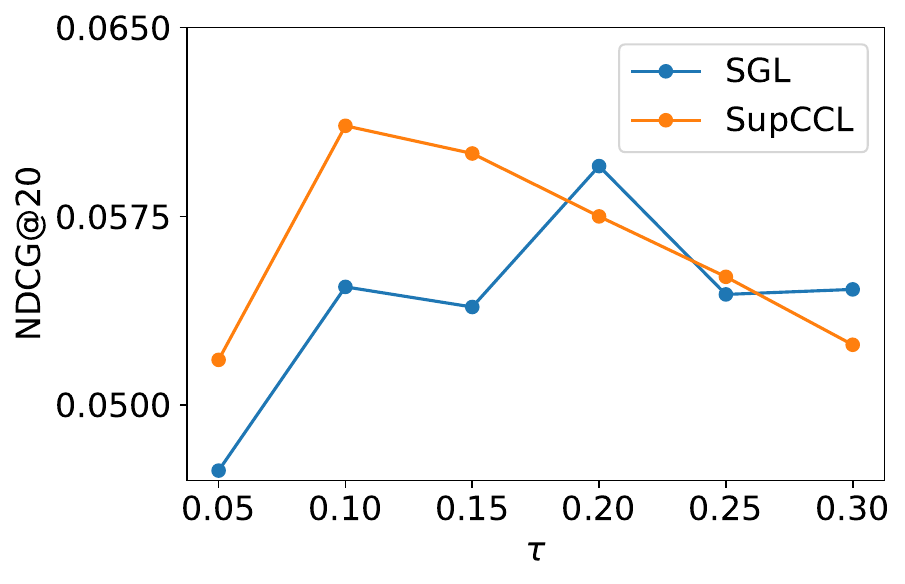}
    \end{minipage}
}
\subfloat[NDCG@20 on Gowalla]{
    \begin{minipage}[b]{0.3\textwidth}
    \includegraphics[width=1\textwidth]{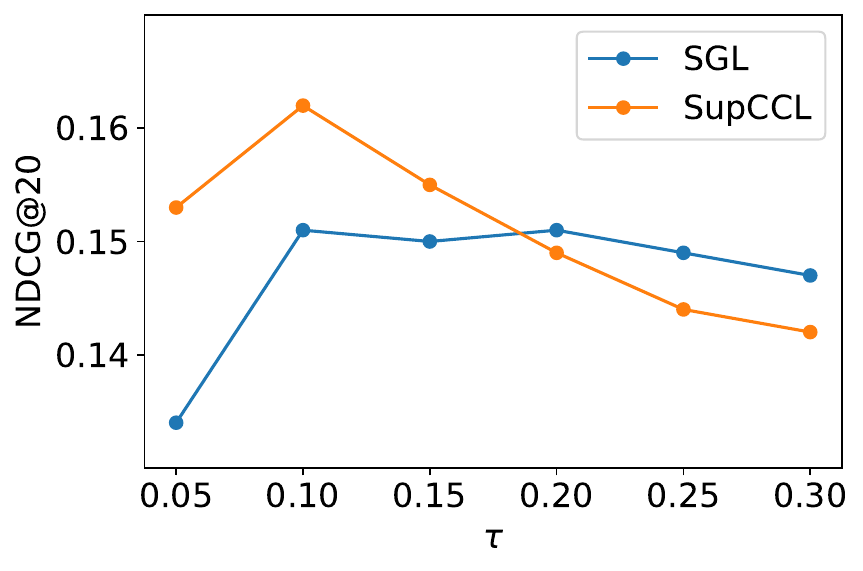}
    \end{minipage}
}
\subfloat[NDCG@20 on Amazon-Book]{
    \begin{minipage}[b]{0.3\textwidth}
    \includegraphics[width=1\textwidth]{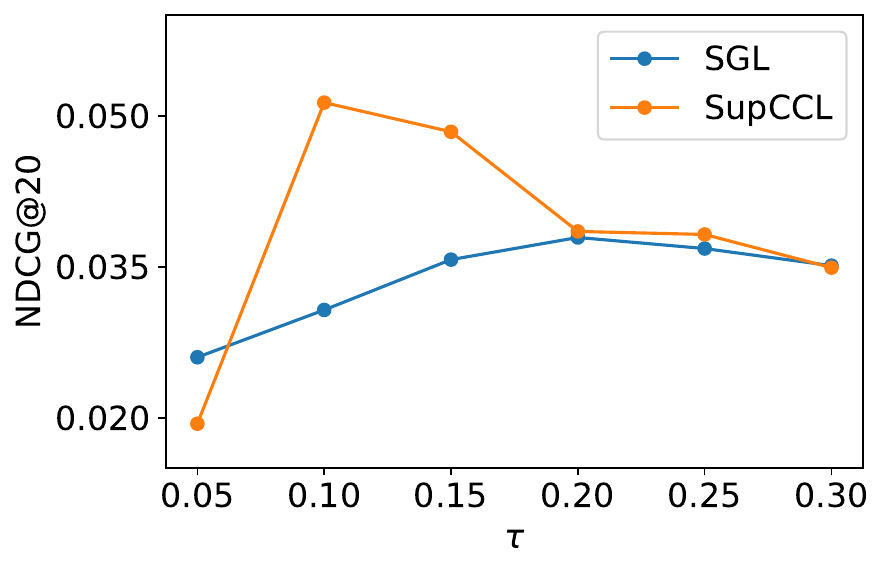}
    \end{minipage}
}
\quad
\caption{Recall@20 and NDCG@20 on three real-world datasets with different temperature $\tau$ values.} \label{fig:ssl_temp}
\end{figure*}

\subsubsection{How the Temperature Values $\tau$ Influence the Performance of Supervised Contrastive Loss}

In this section, we aim to test the performance of our proposed loss functions under different values of temperature $\tau$. 
The overall loss function for our proposed loss functions we use is Equation~\eqref{eq:NESCL_overall_loss}. 
To make the comparison fair, we use the \textit{Edge Drop} augmentation strategy to generate the augmented graphs for both models. The experiments can refer to Figure~\ref{fig:ssl_temp}. 
From the experimental results, we have the following observations. 

1. We test the performance of our proposed with different temperature values $\tau$; we mainly select the candidate values from the list [0.05,0.10,0.15,0.20,0.25,0.30]. From the results, we can find on both metrics Recall@20 and NDCG@20, our proposed loss functions increases first, then drops with the increasing of the $\tau$ values. And our proposed loss functions achieve the best performance when setting the temperature $\tau$ to a small value of 0.1 on all datasets. The $\tau$ is also set to 0.1 in other experiments. 
When the temperature is set to a small value, it can better utilize the information from the negative samples which are not hard. 
Despite the presence of false negative samples, using some of the user’s nearest neighbors as positive samples for contrastive learning enables us to leverage the advantages of smaller temperature coefficients, thereby offsetting the potential adverse impact of these false samples. 
However, when the temperature is set to a value that is smaller than 0.1, the performance of the backbone model based on our proposed loss function drops, which may be because of the gradient explosion issue. 

2. We can find the performance of our proposed loss function degrades with the increase of the temperature $\tau$. We suppose the possible reason is that when the temperature is a large value, as the negative samples are too many, it may weaken the role of the positive samples. Let's take the value $\lambda^{in}_k$ in the subsection~\ref{sec:analysis_loss_in_NESCL} as an example when the temperature is larger, no matter the positive sample $k$ is close to $i$ or apart from $i$, the value which is provided by the term $exp(\mathbf{h}^{'}_k(\mathbf{h}^{''}_i)^{\top}/\tau)$ would be a smaller one, which means the signal which is provided by the $exp(\mathbf{h}^{'}_k(\mathbf{h}^{''}_i)^{\top}/\tau)$ would be more weaken. Thus, the performance of the backbone would be decreased. 

3. From the Equation~\eqref{eq:NESCL_multiple_positives_in} and Equation~\eqref{eq:NESCL_multiple_positives_out}, it is obvious that the gradient of our proposed loss function could be larger than SGL. From the equation of calculating $\lambda^{in}_k$, we can find when the temperature is larger, the negative effects of the negative samples would be enlarged, and the positive effects of the positive samples would be weakened. As our proposed loss function provides two more gradient terms than SGL, it would further exacerbate the above negative effects when the temperature $\tau$ is larger. 
It also supports why our proposed loss functions may be inferior to SGL when the temperature values are larger. 

\begin{table*}[]
\centering
\caption{The Performance of Our Proposed Loss Functions \shortname~under Different Kinds of Nearest Neighbors Incorporating Strategies.}
\label{tab:different_incorporating_strategies}
\begin{tabular}{|ll|ll|ll|ll|}
\hline
\multicolumn{2}{|l|}{\multirow{2}{*}{Experimental Setting}} & \multicolumn{2}{c|}{Yelp2018}            & \multicolumn{2}{c|}{Gowalla}             & \multicolumn{2}{c|}{Amazon-Book}         \\ \cline{3-8} 
\multicolumn{2}{|l|}{}                                      & \multicolumn{1}{l|}{Recall@20} & NDCG@20 & \multicolumn{1}{l|}{Recall@20} & NDCG@20 & \multicolumn{1}{l|}{Recall@20} & NDCG@20 \\ \hline
\multicolumn{2}{|c|}{Identify Weights}                                      & \multicolumn{1}{l|}{0.0725}    & \multicolumn{1}{l|}{0.0596}  & \multicolumn{1}{l|}{0.1891}    & \multicolumn{1}{l|}{0.1603}  & \multicolumn{1}{l|}{0.0567}    & \multicolumn{1}{l|}{0.0473}  \\ \hline
\multicolumn{2}{|c|}{Similarity Weights}                                      & \multicolumn{1}{l|}{0.0724}    & \multicolumn{1}{l|}{0.0595}  & \multicolumn{1}{l|}{0.1894}    & \multicolumn{1}{l|}{0.1603}  & \multicolumn{1}{l|}{0.0562}    & \multicolumn{1}{l|}{0.0470}  \\ \hline
\multicolumn{2}{|c|}{Random Sampling}                                      & \multicolumn{1}{l|}{0.0743}    & \multicolumn{1}{l|}{0.0611}  & \multicolumn{1}{l|}{0.1917}    & \multicolumn{1}{l|}{0.1617}  & \multicolumn{1}{l|}{0.0622}    & \multicolumn{1}{l|}{0.0509}  \\ \hline
\multicolumn{2}{|c|}{Weighted Sampling}                                      & \multicolumn{1}{l|}{0.0739}    & \multicolumn{1}{l|}{0.0609}  & \multicolumn{1}{l|}{0.1914}    & \multicolumn{1}{l|}{0.1616}  & \multicolumn{1}{l|}{0.0624}    & \multicolumn{1}{l|}{0.0513}  \\ \hline
\end{tabular}
\end{table*}
\subsubsection{The Influence of Different Kinds of Collaborative Neighbors Incorporating Strategies.}
\label{sec:colla_neighbors}
We test the influence of the number of nearest neighbors $K$. The set of values for $K$ is (5, 10, 15). From the experiments, we can find that our proposed loss functions perform best when $K$ is set to 15, 5, and 5, for Yelp2018, Gowalla, and Amazon-Book datasets, respectively.  
As we argued that the nearest neighbors are found by the memory-based methods, their quality may not be guaranteed. Thus, in this section, we propose several strategies to incorporate collaborative neighbors. 
First, as the similarity value $sim(\cdot)$ in Equation~\eqref{eq:NESCL_multiple_positives_in} and Equation~\eqref{eq:NESCL_multiple_positives_out} between the nearest neighbors and the anchor node is calculated by the memory-based methods, we would like to test the performance of our proposed loss functions under different similarity settings, such as we treat the $sim(a,i)$ as 1 or the values which are calculated by the memory-based methods. They correspond to the ``Identify Weights'' and ``Similarity Weights'' in Table~\ref{tab:different_incorporating_strategies}. 

Second, as we analyzed in section~\ref{sec:analysis_gradient}, the weights of different kinds of positive samples are influenced by other positive samples in the supervised contrastive loss, and the quality of the nearest neighbors is not guaranteed, incorporating all nearest neighbors may harm the performance of the backbone model. We think it may be more reasonable to randomly select one nearest neighbor in the designed supervised contrastive loss. We propose two kinds of sampling strategies, one is random sampling, which is ``Random Sampling'' in Table~\ref{tab:different_incorporating_strategies}. And another is sampling according to the $sim(a,i)$, which corresponds the ``Weighed Sampling'' in Table~\ref{tab:different_incorporating_strategies}. For the nearest neighbor with a larger similarity value, which may be sampled with a higher probability.  

From the results in Table~\ref{tab:different_incorporating_strategies}, we find the ``Random Sampling'' method achieves the best performance. The results show that our proposed similarity incorporating strategies don't work. We suppose that there are two possible reasons. One is the similarity values are not accurate enough. We think more advanced memory-based methods can be used to get more accurate similarity values. Another one is our proposed similarity-incorporating is ineffective. And we leave it as future work. 
\section{Conclusion and Future Work}

We've developed an effective, supervised, collaborative contrastive loss function, \shortname~. Based on the contrastive loss function, it leverages all positive samples to optimize model performance. Our theoretical analysis reveals that our loss function better adjusts the influence of different positive samples on anchor node representation. Experimental results demonstrate an improved performance over the SGL model on three practical datasets, showing improvements of 10.09\%, 7.09\%, and 35.36\% on the NDCG@20 metric. Furthermore, we've investigated the effect of varying temperature values $\tau$, finding that smaller values result in better performance for the backbone model.

While our work has achieved positive outcomes, several challenges remain unresolved, and opportunities for exploration exist. First, we aim better to incorporate the similarity values between neighboring and anchor nodes to enhance the performance of our proposed supervised contrastive loss function. Second, we've tested our loss function within GNN-based models and aspire to explore its application to other types of input data, such as item sequences and social relation graphs. Lastly, considering the memory consumption issue of our model, we plan to investigate more efficient graph augmentation techniques to address this problem.

%

\appendices

\section{Analysis of the Neighborhood-Enhanced Supervised Contrastive Loss from the Gradient Perspective.}
In this section, to analyze why our proposed supervised collaborative contrastive loss can weigh the importance of different kinds of positive samples. We aim to analyze the loss function from the gradient perspective. In this section, we would provide more details about how to calculate the gradient or $\mathbf{h}''_i$ from loss functions $\mathcal{L}^{in}_{NESCL}$ and $\mathcal{L}^{out}_{NESCL}$. And the analysis of how our proposed loss functions work can refer to sections 5.1 and 5.2 in the original manuscript.

\subsection{Calculating the Gradient From $\mathcal{L}^{in}_{NESCL}$}
When calculating the gradient to $\mathbf{h}''_i$ from $\mathcal{L}^{in}_{NESCL}$, we could have
\begin{align*}
    \frac{\partial\mathcal{L}^{in}_{NESCL}}{\partial\mathbf{h}''_i} =&  \frac{\partial}{\partial\mathbf{h}''_i} (-(log\frac{exp(\mathbf{h}^{'}_i(\mathbf{h}^{''}_i)^{\top}/\tau)}{\sum_{j\in\mathcal{N}}exp(\mathbf{h}^{'}_i(\mathbf{h}^{''}_j)^{\top}/\tau)}
    \\
    &+ log\frac{exp(\mathbf{h}^{'}_{k}(\mathbf{h}^{''}_i)^{\top}/\tau)}{\sum_{j\in\mathcal{N}}exp(\mathbf{h}^{'}_{k}(\mathbf{h}^{''}_j)^{\top}/\tau)}
    \\
    &+ log\frac{exp(\mathbf{h}^{'}_{a}(\mathbf{h}^{''}_i)^{\top}/\tau)}{\sum_{j\in\mathcal{N}}exp(\mathbf{h}^{'}_{a}(\mathbf{h}^{''}_j)^{\top}/\tau)}
    )) 
\end{align*}
to simplify this equation, we define the following notations:
\begin{align*}
    X_1 = \sum_{j\in\mathcal{N}}exp(\mathbf{h}^{'}_i(\mathbf{h}^{''}_j)^{\top}/\tau),
\end{align*}

\begin{align*}
    X_2 = \sum_{j\in\mathcal{N}}exp(\mathbf{h}^{'}_{k}(\mathbf{h}^{''}_j)^{\top}/\tau),
\end{align*}

and 
\begin{align*}
    X_3 = \sum_{j\in\mathcal{N}}exp(\mathbf{h}^{'}_{a}(\mathbf{h}^{''}_j)^{\top}/\tau).
\end{align*}

Then, we can get the following equation:
\begin{align*}
    \frac{\partial\mathcal{L}^{in}_{NESCL}}{\partial\mathbf{h}''_i}
    =& \frac{\partial}{\partial\mathbf{h}''_i}(-(log\frac{exp(\mathbf{h}^{'}_i(\mathbf{h}^{''}_i)^{\top}/\tau)}{X_1}
    \\
    &+log\frac{exp(\mathbf{h}^{'}_{k}(\mathbf{h}^{''}_i)^{\top}/\tau)}{X_2}
    \\
    &+log\frac{exp(\mathbf{h}^{'}_{a}(\mathbf{h}^{''}_i)^{\top}/\tau}{X_3}))
    \\
    =& \frac{\partial}{\partial\mathbf{h}''_i}(log X_1-log(exp(\mathbf{h}^{'}_i(\mathbf{h}^{''}_i)^{\top}/\tau))
    \\
    &+log X_2-log(exp(\mathbf{h}^{'}_{k}(\mathbf{h}^{''}_i)^{\top}/\tau))
    \\
    &+log X_3-log(exp(\mathbf{h}^{'}_{a}(\mathbf{h}^{''}_i)^{\top}/\tau))
    )
    \\
    =& (\frac{\frac{\partial X_1}{\partial\mathbf{h}''_i}}{X_1}-\mathbf{h}'_i/\tau+\frac{\frac{\partial X_2}{\partial\mathbf{h}''_i}}{X_2}-\mathbf{h}'_{k}/\tau+\frac{\frac{\partial X_3}{\partial\mathbf{h}''_i}}{X_3}-\mathbf{h}'_{a}/\tau),\numberthis \label{eq:first}
\end{align*}

where 
\begin{align*}
    \frac{\partial X_1}{\partial \mathbf{h}''_i} = \frac{\mathbf{h}'_i}{\tau} exp(\mathbf{h}^{'}_i(\mathbf{h}^{''}_i)^{\top}/\tau),
\end{align*}

\begin{align*}
    \frac{\partial X_2}{\partial \mathbf{h}''_i} = \frac{\mathbf{h}'_{k}}{\tau}exp(\mathbf{h}^{'}_{k}(\mathbf{h}^{''}_i)^{\top}/\tau),
\end{align*}

and 
\begin{align*}
    \frac{\partial X_3}{\partial \mathbf{h}''_i} = \frac{\mathbf{h}'_{a}}{\tau}exp(\mathbf{h}^{'}_{a}(\mathbf{h}^{''}_i)^{\top}/\tau).
\end{align*}

Before further expanding the above equation, we define the $\frac{\partial \mathcal{L}^{in}_{NESCL}}{\partial\mathbf{h}''_i}$ with:
\begin{align*}
    \frac{\partial\mathcal{L}^{in}_{NESCL}}{\partial\mathbf{h}''_i}=\lambda^{in}_i\mathbf{h}'_i + \lambda^{in}_{k}\mathbf{h}'_{k} + \lambda^{in}_{a}\mathbf{h}'_{a}.
\end{align*}

With the equation~\eqref{eq:first} and the gradients of $\frac{\partial X_1}{\partial \mathbf{h}''_i}$, $\frac{\partial X_2}{\partial \mathbf{h}''_i}$ and $\frac{\partial X_3}{\partial \mathbf{h}''_i}$, we could calculate the term $\lambda^{in}_i\mathbf{h}'_i$ with:
\begin{align*}
    \lambda^{in}_i\mathbf{h}'_i &= 
    \frac{\frac{\mathbf{h}'_i}{\tau}exp(\mathbf{h}^{'}_i(\mathbf{h}^{''}_i)^{\top}/\tau)}{X_1} - \frac{\mathbf{h}_i'}{\tau}
    \\
    &= \frac{\mathbf{h}_i'}{\tau}(\frac{exp(\mathbf{h}^{'}_i(\mathbf{h}^{''}_i)^{\top}/\tau)}{X_1}-1)
    \\
    &= \frac{\mathbf{h}_i'}{\tau} (\frac{exp(\mathbf{h}^{'}_i(\mathbf{h}^{''}_i)^{\top}/\tau)-X_1}{X_1}).
\end{align*}

Thus, the value of $\lambda^{in}_i$ can be denoted as:
\begin{align*}
    \lambda^{in}_i = \frac{1}{\tau} (\frac{exp(\mathbf{h}^{'}_i(\mathbf{h}^{''}_i)^{\top}/\tau)-X_1}{X_1}).
\end{align*}

Similarly, the value of $\lambda^{in}_{k}$ and $\lambda^{in}_{a}$ can be denoted as:
\begin{align*}
    \lambda^{in}_{k} = \frac{1}{\tau} (\frac{exp(\mathbf{h}^{'}_{k}(\mathbf{h}^{''}_i)^{\top}/\tau)-X_2}{X_2}),
\end{align*}

and 
\begin{align*}
    \lambda^{in}_{a} = \frac{1}{\tau} (\frac{exp(\mathbf{h}^{'}_{a}(\mathbf{h}^{''}_i)^{\top}/\tau)-X_3}{X_3}).
\end{align*}

By substituting the expansion of the $X_1$ into the formula corresponding to the $\lambda^{in}_i$, we can get the following equation:
\begin{align*}
    \lambda^{in}_i &= \frac{1}{\tau} (\frac{exp(\mathbf{h}^{'}_i(\mathbf{h}^{''}_i)^{\top}/\tau) - \sum_{j\in\mathcal{N}}exp(\mathbf{h}^{'}_i(\mathbf{h}^{''}_j)^{\top}/\tau)}{\sum_{j\in\mathcal{N}}exp(\mathbf{h}^{'}_i(\mathbf{h}^{''}_j)^{\top}/\tau)})
    \\
    &= \frac{1}{\tau} (\frac{-\sum_{j\in\mathcal{N},j\neq i}exp(\mathbf{h}^{'}_i(\mathbf{h}^{''}_j)^{\top}/\tau)}{exp(\mathbf{h}^{'}_i(\mathbf{h}^{''}_i)^{\top}/\tau)+\sum_{j\in\mathcal{N},j\neq i}exp(\mathbf{h}^{'}_i(\mathbf{h}^{''}_j)^{\top}/\tau)}).
\end{align*}

Similarly, we can get the value of the $\lambda^{in}_{k}$ and $\lambda^{in}_{a}$ with:
\begin{align*}
    \lambda^{in}_{k}=\frac{1}{\tau}(\frac{-\sum_{j\in\mathcal{N},j\neq i}exp(\mathbf{h}^{'}_{k}(\mathbf{h}^{''}_j)^{\top}/\tau)}{exp(\mathbf{h}^{'}_{k}(\mathbf{h}^{''}_i)^{\top}/\tau)+\sum_{j\in\mathcal{N},j\neq i}exp(\mathbf{h}^{'}_{k}(\mathbf{h}^{''}_j)^{\top}/\tau)}),
\end{align*}

and 
\begin{align*}
    \lambda^{in}_{a}=\frac{1}{\tau}(\frac{-\sum_{j\in\mathcal{N},j\neq i}exp(\mathbf{h}^{'}_{a}(\mathbf{h}^{''}_j)^{\top}/\tau)}{exp(\mathbf{h}^{'}_{a}(\mathbf{h}^{''}_i)^{\top}/\tau)+\sum_{j\in\mathcal{N},j\neq i}exp(\mathbf{h}^{'}_{a}(\mathbf{h}^{''}_j)^{\top}/\tau)}).
\end{align*}

The analysis of how these values can be used to identify the importance of different positive samples can refer to section 5.1. 

\subsection{Calculating the Gradient From $\mathcal{L}^{out}_{NESCL}$}
When calculating the gradient from $\mathcal{L}^{out}_{NESCL}$ to $\mathbf{h}''_i$, we have
\begin{align*}
    \frac{\partial\mathcal{L}^{out}_{NESCL}}{\partial\mathbf{h}''_i} 
    =& 
    \frac{\partial}{\partial\mathbf{h}''_i} (
    -log(\frac{exp(\mathbf{h}^{'}_i(\mathbf{h}^{''}_i)^{\top}/\tau)}{\sum_{j\in\mathcal{N}}exp(\mathbf{h}^{'}_i(\mathbf{h}^{''}_j)^{\top}/\tau)}
    \\
    &+ \frac{exp(\mathbf{h}^{'}_{k}(\mathbf{h}^{''}_i)^{\top}/\tau)}{\sum_{j\in\mathcal{N}}exp(\mathbf{h}^{'}_{k}(\mathbf{h}^{''}_j)^{\top}/\tau)}
    \\
    &+ \frac{exp(\mathbf{h}^{'}_{a}(\mathbf{h}^{''}_i)^{\top}/\tau)}{\sum_{j\in\mathcal{N}}exp(\mathbf{h}^{'}_{a}(\mathbf{h}^{''}_j)^{\top}/\tau)}
    )), 
\end{align*}
to simplify this equation with $X_1$, $X_2$, and $X_3$, we can get following equation:
\begin{align*}
    \frac{\partial\mathcal{L}^{out}_{NESCL}}{\partial\mathbf{h}''_i} =& 
    \frac{\partial}{\partial\mathbf{h}''_i}(-log(\frac{exp(\mathbf{h}^{'}_i(\mathbf{h}^{''}_i)^{\top}/\tau)}{X_1}
    \\
    &+ \frac{exp(\mathbf{h}^{'}_{k}(\mathbf{h}^{''}_i)^{\top}/\tau)}{X_2} 
    \\
    &+ \frac{exp(\mathbf{h}^{'}_{a}(\mathbf{h}^{''}_i)^{\top}/\tau)}{X_3}
    ))
    \\
    =& \frac{\partial}{\partial\mathbf{h}''_i}(log(X_1X_2X_3)
    \\
    &-log(exp(\mathbf{h}^{'}_i(\mathbf{h}^{''}_i)^{\top}/\tau)X_2X_3
    \\
    &+exp(\mathbf{h}^{'}_{k}(\mathbf{h}^{''}_i)^{\top}/\tau)X_1X_3
    \\
    &+exp(\mathbf{h}^{'}_{a}(\mathbf{h}^{''}_i)^{\top}/\tau)X_1X_2)). \numberthis \label{eq:second}
\end{align*}

To further simplify this equation, we let
\begin{align*}
    &Y=exp(\mathbf{h}^{'}_i(\mathbf{h}^{''}_i)^{\top}/\tau)X_2X_3+exp(\mathbf{h}^{'}_{k}(\mathbf{h}^{''}_i)^{\top}/\tau)X_1X_3
    \\
    &+exp(\mathbf{h}^{'}_{a}(\mathbf{h}^{''}_i)^{\top}/\tau)X_1X_2.
\end{align*}

And we can simplify above Equation~\eqref{eq:second} to:
\begin{align*}
    \frac{\partial\mathcal{L}^{out}_{NESCL}}{\partial\mathbf{h}''_i} =& \frac{\partial}{\partial\mathbf{h}''_i}(log(X_1X_2X_3)-log(Y))
    \\
    =&(
    \frac{\frac{\partial X_1}{\partial \mathbf{h}''_i}}{X_1}
    +\frac{\frac{\partial X_2}{\partial \mathbf{h}''_i}}{X_2}
    +\frac{\frac{\partial X_3}{\partial \mathbf{h}''_i}}{X_3}
    -\frac{\frac{\partial Y}{\partial\mathbf{h}''_i}}{Y}), \numberthis \label{eq:third}
\end{align*}

where

\begin{align*}
    \frac{\partial Y}{\partial \mathbf{h}''_i} =&
     \frac{\partial}{\partial \mathbf{h}''_i} (exp(\mathbf{h}^{'}_i(\mathbf{h}^{''}_i)^{\top}/\tau)X_2X_3
    \\
    &+exp(\mathbf{h}^{'}_{k}(\mathbf{h}^{''}_i)^{\top}/\tau)X_1X_3
    \\
    &+exp(\mathbf{h}^{'}_{a}(\mathbf{h}^{''}_i)^{\top}/\tau)X_1X_2)
    \\
    =& \frac{\mathbf{h}'_i}{\tau}exp(\mathbf{h}^{'}_i(\mathbf{h}^{''}_i)^{\top}/\tau)X_2X_3 
    \\
    &+ exp(\mathbf{h}^{'}_i(\mathbf{h}^{''}_i)^{\top}/\tau)\frac{\mathbf{h}'_{k}}{\tau}exp(\mathbf{h}^{'}_{k}(\mathbf{h}^{''}_i)^{\top}/\tau)X_3 
    \\
    &+
    exp(\mathbf{h}^{'}_i(\mathbf{h}^{''}_i)^{\top}/\tau)X_2\frac{\mathbf{h}'_{a}}{\tau}exp(\mathbf{h}^{'}_{a}(\mathbf{h}^{''}_i)^{\top}/\tau)
    \\
    &+ \frac{\mathbf{h}'_{k}}{\tau}exp(\mathbf{h}^{'}_{k}(\mathbf{h}^{''}_i)^{\top}/\tau)X_1X_3
    \\
    &+ exp(\mathbf{h}^{'}_{k}(\mathbf{h}^{''}_i)^{\top}/\tau) \frac{\mathbf{h}'_i}{\tau} exp(\mathbf{h}^{'}_i(\mathbf{h}^{''}_i)^{\top}/\tau) X_3
    \\
    &+ exp(\mathbf{h}^{'}_{k}(\mathbf{h}^{''}_i)^{\top}/\tau)X_1\frac{\mathbf{h}'_{a}}{\tau}exp(\mathbf{h}^{'}_a(\mathbf{h}^{''}_i)^{\top}/\tau)
    \\
    &+ \frac{\mathbf{h}_{a}'}{\tau}exp(\mathbf{h}^{'}_{a}(\mathbf{h}^{''}_i)^{\top}/\tau)X_1X_2
    \\
    &+
    exp(\mathbf{h}^{'}_{a}(\mathbf{h}^{''}_i)^{\top}/\tau)\frac{\mathbf{h}'_i}{\tau} exp(\mathbf{h}^{'}_i(\mathbf{h}^{''}_i)^{\top}/\tau) X_2
    +
    \\
    &+
    exp(\mathbf{h}^{'}_{a}(\mathbf{h}^{''}_i)^{\top}/\tau)X_1\frac{\mathbf{h}'_{k}}{\tau}exp(\mathbf{h}^{'}_{k}(\mathbf{h}^{''}_i)^{\top}/\tau)
    .
\end{align*}

Before further expanding Equation~\eqref{eq:third}, we define the $\frac{\partial\mathcal{L}^{out}_{NESCL}}{\partial\mathbf{h}''_i}$ as:
\begin{equation}
    \frac{\partial\mathcal{L}^{out}_{NESCL}}{\partial\mathbf{h}''_i}=\lambda^{out}_i\mathbf{h}'_i + \lambda^{out}_{k}\mathbf{h}'_{k} + \lambda^{out}_{a}\mathbf{h}'_{a}.
\end{equation}

With Equation~\eqref{eq:third} and the gradient values of $\frac{\partial X_1}{\mathbf{h}''_i}$, $\frac{\partial X_2}{\mathbf{h}''_i}$, $\frac{\partial X_3}{\mathbf{h}''_i}$, and $\frac{\partial Y}{\mathbf{h}''_i}$, we could calculate the term $\lambda^{out}_i \mathbf{h}'_i$ with:
\begin{align*}
    \lambda^{out}_i\mathbf{h}'_i
    =&
    \frac{\frac{\mathbf{h}'_i}{\tau} exp(\mathbf{h}^{'}_i(\mathbf{h}^{''}_i)^{\top}/\tau)}{X_1}
    \\
    &-\frac{\frac{\mathbf{h}'_i}{\tau}exp(\mathbf{h}^{'}_i(\mathbf{h}^{''}_i)^{\top}/\tau)X_2X_3}{Y}
    \\
    &-\frac{exp(\mathbf{h}^{'}_{k}(\mathbf{h}^{''}_i)^{\top}/\tau) \frac{\mathbf{h}'_i}{\tau} exp(\mathbf{h}^{'}_i(\mathbf{h}^{''}_i)^{\top}/\tau) X_3}{Y}
    \\
    &-\frac{exp(\mathbf{h}^{'}_{a}(\mathbf{h}^{''}_i)^{\top}/\tau)\frac{\mathbf{h}'_i}{\tau} exp(\mathbf{h}^{'}_i(\mathbf{h}^{''}_i)^{\top}/\tau) X_2}{Y}
    \\
    &=
    \frac{\frac{\mathbf{h}'_i}{\tau} exp(\mathbf{h}^{'}_i(\mathbf{h}^{''}_i)^{\top}/\tau)Y}{X_1Y}
    \\
    &-\frac{X_1\frac{\mathbf{h}'_i}{\tau}exp(\mathbf{h}^{'}_i(\mathbf{h}^{''}_i)^{\top}/\tau)X_2X_3}{X_1Y}
    \\
    &-\frac{X_1exp(\mathbf{h}^{'}_{k}(\mathbf{h}^{''}_i)^{\top}/\tau) \frac{\mathbf{h}'_i}{\tau} exp(\mathbf{h}^{'}_i(\mathbf{h}^{''}_i)^{\top}/\tau) X_3}{X_1Y}
    \\
    &-\frac{X_1exp(\mathbf{h}^{'}_{a}(\mathbf{h}^{''}_i)^{\top}/\tau)\frac{\mathbf{h}'_i}{\tau} exp(\mathbf{h}^{'}_i(\mathbf{h}^{''}_i)^{\top}/\tau) X_2}{X_1Y}
    \\
    =&
    \frac{\mathbf{h}_i'}{\tau}\frac{exp(\mathbf{h}^{'}_i(\mathbf{h}^{''}_i)^{\top}/\tau)-X_1}{X_1}\frac{exp(\mathbf{h}^{'}_i(\mathbf{h}^{''}_i)^{\top}/\tau)X_2X_3}{Y}.
\end{align*}
Thus, the value of $\lambda^{out}_i$ is:
\begin{equation}
    \lambda^{out}_i = \frac{1}{\tau}\frac{exp(\mathbf{h}^{'}_i(\mathbf{h}^{''}_i)^{\top}/\tau)-X_1}{X_1}\frac{exp(\mathbf{h}^{'}_i(\mathbf{h}^{''}_i)^{\top}/\tau)X_2X_3}{Y}. 
\end{equation}

Also, we can get:
\begin{align*}
    \lambda^{out}_{k} = \frac{1}{\tau}\frac{exp(\mathbf{h}^{'}_{k}(\mathbf{h}^{''}_i)^{\top}/\tau)-X_2}{X_2}\frac{exp(\mathbf{h}^{'}_{k}(\mathbf{h}^{''}_i)^{\top}/\tau)X_1X_3}{Y},
\end{align*}
and 
\begin{align*}
    \lambda^{out}_{a} = \frac{1}{\tau}\frac{exp(\mathbf{h}^{'}_{a}(\mathbf{h}^{''}_i)^{\top}/\tau)-X_3}{X_3}\frac{exp(\mathbf{h}^{'}_{a}(\mathbf{h}^{''}_i)^{\top}/\tau)X_1X_2}{Y}.
\end{align*}

From the above analysis, it is not easy to get any observations. As the formula of $\lambda^{out}_i$ is quite complex, we divide $\lambda^{out}_i$ by $\lambda^{in}_i$, and we can get:
\begin{align*}
    \frac{\lambda^{out}_i}{\lambda^{in}_i}=&
    \frac{Y}{exp(\mathbf{h}^{'}_i(\mathbf{h}^{''}_i)^{\top}/\tau)X_2X_3}
    \\
    =&\frac{exp(\mathbf{h}^{'}_i(\mathbf{h}^{''}_i)^{\top}/\tau)X_2X_3}{exp(\mathbf{h}^{'}_i(\mathbf{h}^{''}_i)^{\top}/\tau)X_2X_3}
    +\frac{exp(\mathbf{h}^{'}_{k}(\mathbf{h}^{''}_i)^{\top}/\tau)X_1X_3}{exp(\mathbf{h}^{'}_i(\mathbf{h}^{''}_i)^{\top}/\tau)X_2X_3}
    \\
    &+\frac{exp(\mathbf{h}^{'}_{a}(\mathbf{h}^{''}_i)^{\top}/\tau)X_1X_2}{exp(\mathbf{h}^{'}_i(\mathbf{h}^{''}_i)^{\top}/\tau)X_2X_3}
    \\
    =&1+\frac{exp(\mathbf{h}^{'}_{k}(\mathbf{h}^{''}_i)^{\top}/\tau)X_1}{exp(\mathbf{h}^{'}_i(\mathbf{h}^{''}_i)^{\top}/\tau)X_2}+\frac{exp(\mathbf{h}^{'}_{a}(\mathbf{h}^{''}_i)^{\top}/\tau)X_1}{exp(\mathbf{h}^{'}_i(\mathbf{h}^{''}_i)^{\top}/\tau)X_3}
    \\
    =&
    1+\frac{1+\frac{\sum_{j\in\mathcal{N},j\neq i}exp(\mathbf{h}^{'}_i(\mathbf{h}^{''}_j)^{\top}/\tau)}{exp(\mathbf{h}^{'}_i(\mathbf{h}^{''}_i)^{\top}/\tau)}}{1+\frac{\sum_{j\in\mathcal{N},j\neq i}exp(\mathbf{h}^{'}_{k}(\mathbf{h}^{''}_j)^{\top}/\tau)}{exp(\mathbf{h}^{'}_{k}(\mathbf{h}^{''}_i)^{\top}/\tau)}}
    \\
    +&
    \frac{1+\frac{\sum_{j\in\mathcal{N},j\neq i}exp(\mathbf{h}^{'}_i(\mathbf{h}^{''}_j)^{\top}/\tau)}{exp(\mathbf{h}^{'}_i(\mathbf{h}^{''}_i)^{\top}/\tau)}}{1+\frac{\sum_{j\in\mathcal{N},j\neq i}exp(\mathbf{h}^{'}_{a}(\mathbf{h}^{''}_j)^{\top}/\tau)}{exp(\mathbf{h}^{'}_{a}(\mathbf{h}^{''}_i)^{\top}/\tau)}}
\end{align*}

The analysis of how these values can be used to identify the importance of different positive samples can refer to section 5.2.

\section{Experimental Results}
In the original manuscript, we have reported the performance of the backbone model based on our proposed loss function under some key settings, such as different combinations of loss functions, temperature values, and sampling strategies of the nearest neighbors. 
While in this section, we will show how different hyperparameters influence the performance of the backbone model based on our proposed loss function, such as data augmentation strategy, data augmentation ratio $\rho$, layer number, and coefficient value $\alpha$ in the GNN model, which are not very important to our proposed model. 

\begin{table*}[]
\centering
\caption{The Performance of Our Proposed Model on Different GNN Layers and Data Augmentation Strategies.}
\label{tab:gnn_layers_data_aug}
\begin{tabular}{|ll|ll|ll|ll|}
\hline
\multicolumn{2}{|c|}{\multirow{2}{*}{Model}}                & \multicolumn{2}{c|}{Yelp2018}            & \multicolumn{2}{c|}{Gowalla}             & \multicolumn{2}{c|}{Amazon-Book}         \\ \cline{3-8} 
\multicolumn{2}{|l|}{}                                      & \multicolumn{1}{l|}{Recall@20} & NDCG@20 & \multicolumn{1}{l|}{Recall@20} & NDCG@20 & \multicolumn{1}{l|}{Recall@20} & NDCG@20 \\ \hline
\multicolumn{1}{|l|}{\multirow{7}{*}{1-Layer}} & LightGCN   & \multicolumn{1}{l|}{0.0631}    & 0.0515  & \multicolumn{1}{l|}{0.1755}    & 0.1492  & \multicolumn{1}{l|}{0.0384}    & 0.0298  \\ \cline{2-8} 
\multicolumn{1}{|l|}{}                         & SGL(ND)     & \multicolumn{1}{l|}{0.0643}    & 0.0529  & \multicolumn{1}{l|}{0.1497}    & 0.1259  & \multicolumn{1}{l|}{0.0432}    & 0.0334  \\ \cline{2-8} 
\multicolumn{1}{|l|}{}                         & SGL(ED)     & \multicolumn{1}{l|}{0.0637}    & 0.0526  & \multicolumn{1}{l|}{0.1718}    & 0.1455  & \multicolumn{1}{l|}{0.0451}    & 0.0353  \\ \cline{2-8} 
\multicolumn{1}{|l|}{}                         & SGL(RW)     & \multicolumn{1}{l|}{0.0637}    & 0.0526  & \multicolumn{1}{l|}{0.1639}    & 0.1388  & \multicolumn{1}{l|}{0.0451}    & 0.0353  \\ \cline{2-8} 
\multicolumn{1}{|l|}{}                         & \shortname(ND) & \multicolumn{1}{l|}{0.0722}    & 0.0595  & \multicolumn{1}{l|}{0.1875}    & 0.1585  & \multicolumn{1}{l|}{0.0559}    & 0.0446  \\ \cline{2-8} 
\multicolumn{1}{|l|}{}                         & \shortname(ED) & \multicolumn{1}{l|}{0.0725}    & 0.0599  & \multicolumn{1}{l|}{0.1855}    & 0.1565  & \multicolumn{1}{l|}{0.0596}    & 0.0486  \\ \cline{2-8} 
\multicolumn{1}{|l|}{}                         & \shortname(RW) & \multicolumn{1}{l|}{0.0724}    & 0.0597  & \multicolumn{1}{l|}{0.1857}    & 0.1569  & \multicolumn{1}{l|}{0.0602}    & 0.0492  \\ \hline
\multicolumn{1}{|l|}{\multirow{7}{*}{2-Layer}} & LightGCN   & \multicolumn{1}{l|}{0.0622}    & 0.0504  & \multicolumn{1}{l|}{0.1777}    & 0.1524  & \multicolumn{1}{l|}{0.0411}    & 0.0315  \\ \cline{2-8} 
\multicolumn{1}{|l|}{}                         & SGL(ND)     & \multicolumn{1}{l|}{0.0658}    & 0.0538  & \multicolumn{1}{l|}{0.1656}    & 0.1393  & \multicolumn{1}{l|}{0.0427}    & 0.0335  \\ \cline{2-8} 
\multicolumn{1}{|l|}{}                         & SGL(ED)     & \multicolumn{1}{l|}{0.0668}    & 0.0549  & \multicolumn{1}{l|}{0.1763}    & 0.1492  & \multicolumn{1}{l|}{0.0468}    & 0.0371  \\ \cline{2-8} 
\multicolumn{1}{|l|}{}                         & SGL(RW)     & \multicolumn{1}{l|}{0.0644}    & 0.0530  & \multicolumn{1}{l|}{0.1729}    & 0.1470  & \multicolumn{1}{l|}{0.0453}    & 0.0358  \\ \cline{2-8} 
\multicolumn{1}{|l|}{}                         & \shortname(ND) & \multicolumn{1}{l|}{\textbf{0.0743}}    & \textbf{0.0611}  & \multicolumn{1}{l|}{0.1903}    & 0.1609  & \multicolumn{1}{l|}{0.0548}    & 0.0436  \\ \cline{2-8} 
\multicolumn{1}{|l|}{}                         & \shortname(ED) & \multicolumn{1}{l|}{0.0735}    & 0.0608  & \multicolumn{1}{l|}{0.1855}    & 0.1575  & \multicolumn{1}{l|}{\textbf{0.0624}}    & \textbf{0.0513}  \\ \cline{2-8} 
\multicolumn{1}{|l|}{}                         & \shortname(RW) & \multicolumn{1}{l|}{0.0739}    & 0.0609  & \multicolumn{1}{l|}{0.1871}    & 0.1583  & \multicolumn{1}{l|}{0.0621}    & 0.0512  \\ \hline
\multicolumn{1}{|l|}{\multirow{7}{*}{3-Layer}} & LightGCN   & \multicolumn{1}{l|}{0.0639}    & 0.0525  & \multicolumn{1}{l|}{0.1823}    & 0.1555  & \multicolumn{1}{l|}{0.0410}    & 0.0318  \\ \cline{2-8} 
\multicolumn{1}{|l|}{}                         & SGL(ND)     & \multicolumn{1}{l|}{0.0644}    & 0.0528  & \multicolumn{1}{l|}{0.1719}    & 0.1450  & \multicolumn{1}{l|}{0.0440}    & 0.0346  \\ \cline{2-8} 
\multicolumn{1}{|l|}{}                         & SGL(ED)     & \multicolumn{1}{l|}{0.0675}    & 0.0555  & \multicolumn{1}{l|}{0.1787}    & 0.1510  & \multicolumn{1}{l|}{0.0478}    & 0.0379  \\ \cline{2-8} 
\multicolumn{1}{|l|}{}                         & SGL(RW)     & \multicolumn{1}{l|}{0.0667}    & 0.0547  & \multicolumn{1}{l|}{0.1777}    & 0.1509  & \multicolumn{1}{l|}{0.0457}    & 0.0356  \\ \cline{2-8} 
\multicolumn{1}{|l|}{}                         & \shortname(ND) & \multicolumn{1}{l|}{0.0737}    & 0.0606  & \multicolumn{1}{l|}{\textbf{0.1917}}    & \textbf{0.1617}  & \multicolumn{1}{l|}{0.0542}    & 0.0432  \\ \cline{2-8} 
\multicolumn{1}{|l|}{}                         & \shortname(ED) & \multicolumn{1}{l|}{0.0738}    & 0.0608  & \multicolumn{1}{l|}{0.1897}    & 0.1605  & \multicolumn{1}{l|}{0.0575}    & 0.0469  \\ \cline{2-8} 
\multicolumn{1}{|l|}{}                         & \shortname(RW) & \multicolumn{1}{l|}{0.0736}    & 0.0606  & \multicolumn{1}{l|}{0.1879}    & 0.1593  & \multicolumn{1}{l|}{0.0607}    & 0.0499  \\ \hline
\end{tabular}
\end{table*}
\subsection{The Influence of Different Number of GNN Layers and Different Kinds of Data Augmentation Strategies.}
We test the performance of our proposed model \shortname~ over different GNN layers and different kinds of data augmentation strategies in Table~\ref{tab:gnn_layers_data_aug}. For any number of GNN layers, and data augmentation strategy, our proposed model \shortname~ outperforms the LightGCN and SGL among all datasets. For Yelp2018, when setting GNN layer to 2, and data augmentation strategy to node dropout, our proposed model performs beset. For Gowalla, GNN layer is set to 3, the data augmentation type is set to node dropout. And for Amazon-Book, the GNN layer is set to 2, and we adopt the edge dropout as the data augmentation strategy.

\begin{table*}[]
\centering
\caption{The Performance of Our Proposed Model over Different Data Augmentation Ratios.}
\label{tab:ratios}
\begin{tabular}{|ll|ll|ll|ll|}
\hline
\multicolumn{2}{|l|}{\multirow{2}{*}{Model}}                    & \multicolumn{2}{l|}{Yelp2018}            & \multicolumn{2}{l|}{Gowalla}             & \multicolumn{2}{l|}{Amazon-Book}         \\ \cline{3-8} 
\multicolumn{2}{|l|}{}                                          & \multicolumn{1}{l|}{NDCG@20} & Recall@20 & \multicolumn{1}{l|}{Recall@20} & NDCG@20 & \multicolumn{1}{l|}{Recall@20} & NDCG@20 \\ \hline
\multicolumn{1}{|l|}{\multirow{2}{*}{$\rho$=0.1}} & SGL(ED)  & \multicolumn{1}{l|}{0.0686}  & 0.0563    & \multicolumn{1}{l|}{0.1771}    & 0.1498  & \multicolumn{1}{l|}{0.0451}    & 0.0353  \\ \cline{2-8} 
\multicolumn{1}{|l|}{}                                & \shortname & \multicolumn{1}{l|}{0.0735}  & 0.0605    & \multicolumn{1}{l|}{0.1903}    & 0.1611  & \multicolumn{1}{l|}{0.0616}    & 0.0507  \\ \hline
\multicolumn{1}{|l|}{\multirow{2}{*}{$\rho$=0.3}} & SGL(ED)  & \multicolumn{1}{l|}{0.0680}  & 0.0560    & \multicolumn{1}{l|}{0.1779}    & 0.1510  & \multicolumn{1}{l|}{0.0480}    & 0.0377  \\ \cline{2-8} 
\multicolumn{1}{|l|}{}                                & \shortname & \multicolumn{1}{l|}{\textbf{0.0743}}  & \textbf{0.0611}    & \multicolumn{1}{l|}{\textbf{0.1913}}    & \textbf{0.1617}  & \multicolumn{1}{l|}{\textbf{0.0624}}    & \textbf{0.0513}  \\ \hline
\multicolumn{1}{|l|}{\multirow{2}{*}{$\rho$=0.5}} & SGL(ED)  & \multicolumn{1}{l|}{0.0690}  & 0.0595    & \multicolumn{1}{l|}{0.1781}    & 0.1511  & \multicolumn{1}{l|}{0.0469}    & 0.0370  \\ \cline{2-8} 
\multicolumn{1}{|l|}{}                                & \shortname & \multicolumn{1}{l|}{0.0736}  & 0.0608    & \multicolumn{1}{l|}{0.1891}    & 0.1605  & \multicolumn{1}{l|}{0.0623}    & 0.0512  \\ \hline
\multicolumn{1}{|l|}{\multirow{2}{*}{$\rho$=0.7}} & SGL(ED)  & \multicolumn{1}{l|}{0.0685}  & 0.0565    & \multicolumn{1}{l|}{0.1787}    & 0.1510  & \multicolumn{1}{l|}{0.0469}    & 0.0366  \\ \cline{2-8} 
\multicolumn{1}{|l|}{}                                & \shortname & \multicolumn{1}{l|}{0.0721}  & 0.0595    & \multicolumn{1}{l|}{0.1852}    & 0.1571  & \multicolumn{1}{l|}{0.0622}    & 0.0511  \\ \hline
\multicolumn{1}{|l|}{\multirow{2}{*}{$\rho$=0.9}} & SGL(ED)  & \multicolumn{1}{l|}{0.0626}  & 0.0513    & \multicolumn{1}{l|}{0.1480}    & 0.1164  & \multicolumn{1}{l|}{0.0455}    & 0.0364  \\ \cline{2-8} 
\multicolumn{1}{|l|}{}                                & \shortname & \multicolumn{1}{l|}{0.0651}  & 0.0539    & \multicolumn{1}{l|}{0.1741}    & 0.1485  & \multicolumn{1}{l|}{0.0601}    & 0.0487  \\ \hline
\end{tabular}
\end{table*}

\subsection{The Influence of the Data Augmentation Ratio $\rho$.}
We show the performance of our proposed model \shortname~ and SGL over different data augmentation ratios in Table~\ref{tab:ratios}. For the Yelp2018 and Gowalla, we adopt the node dropout strategy, while for Amazon-Book, the edge dropout strategy is selected. For all datasets, when setting the $\rho$ to 0.3, our proposed model achieves the best performance. The reason why SGL and our proposed model can still work when the $\rho$ is very large is because 
optimizing the ranking loss $\mathcal{L}^{in}_O$ or $\mathcal{L}^{out}_O$ in Equation (10) in the original manuscript works.

\begin{table*}[]
\centering
\caption{The Performance of Our Proposed Model \shortname~ Over Different Coefficient Values $\alpha$.}
\label{tab:regularization}
\begin{tabular}{|ll|ll|ll|ll|}
\hline
\multicolumn{2}{|l|}{\multirow{2}{*}{Model}}                  & \multicolumn{2}{l|}{Yelp2018}            & \multicolumn{2}{l|}{Gowalla}             & \multicolumn{2}{l|}{Amazon-Book}         \\ \cline{3-8} 
\multicolumn{2}{|l|}{}                                        & \multicolumn{1}{l|}{Recall@20} & NDCG@20 & \multicolumn{1}{l|}{Recall@20} & NDCG@20 & \multicolumn{1}{l|}{Recall@20} & NDCG@20 \\ \hline
\multicolumn{1}{|l|}{\multirow{2}{*}{$\alpha$=0.1}} & SGL(ED)  & \multicolumn{1}{l|}{0.0690}    & 0.0595  & \multicolumn{1}{l|}{0.1787}    & 0.1510  & \multicolumn{1}{l|}{0.0451}    & 0.0353  \\ \cline{2-8} 
\multicolumn{1}{|l|}{}                              & \shortname & \multicolumn{1}{l|}{0.0736}    & 0.0608  & \multicolumn{1}{l|}{\textbf{0.1917}}    & \textbf{0.1617}  & \multicolumn{1}{l|}{0.0621}    & 0.0509  \\ \hline
\multicolumn{1}{|l|}{\multirow{2}{*}{$\alpha$=0.3}} & SGL(ED)  & \multicolumn{1}{l|}{0.0686}    & 0.0563  & \multicolumn{1}{l|}{0.1683}    & 0.1417  & \multicolumn{1}{l|}{0.0467}    & 0.0367  \\ \cline{2-8} 
\multicolumn{1}{|l|}{}                              & \shortname & \multicolumn{1}{l|}{\textbf{0.0743}}    & \textbf{0.0611}  & \multicolumn{1}{l|}{0.1872}    & 0.1587  & \multicolumn{1}{l|}{\textbf{0.0624}}    & \textbf{0.0513}  \\ \hline
\multicolumn{1}{|l|}{\multirow{2}{*}{$\alpha$=0.5}} & SGL(ED)  & \multicolumn{1}{l|}{0.0679}    & 0.0599  & \multicolumn{1}{l|}{0.1718}    & 0.1428  & \multicolumn{1}{l|}{0.0478}    & 0.0379  \\ \cline{2-8} 
\multicolumn{1}{|l|}{}                              & \shortname & \multicolumn{1}{l|}{0.0735}    & 0.0608  & \multicolumn{1}{l|}{0.1871}    & 0.1582  & \multicolumn{1}{l|}{0.0622}    & 0.0509  \\ \hline
\end{tabular}
\end{table*}
\subsection{The Performance of Our Proposed Model \shortname~ Over Different Regularization Coefficient Values $\alpha$.}
We conduct experiments on three real-world datasets to show the performance of our proposed model \shortname~ over different regularization coefficient values $\alpha$ in Table ~\ref{tab:regularization}. From the results, we could find our proposed model \shortname~ can achieve the best performance when setting $\alpha$ to 0.3, 0.1, and 0.3 for Yelp2018, Gowalla, and Amazon-Book datasets, respectively.

\ifCLASSOPTIONcompsoc
  \section*{Acknowledgments}
\else
  \section*{Acknowledgment}
\fi

This work was supported in part by grants from the National Natural Science Foundation of China(Grant No. 61972125, U1936219,  61725203, 61732008, 91846201, 62006066), the Open Project Program of the National
Laboratory of Pattern Recognition (NLPR), and 
the Fundamental Research Funds for the  Central Universities (Grant No. JZ2021HGTB0075). Le Wu greatly thanks the support of Young Elite Scientists Sponsorship. And Peijie Sun greatly thanks the support of the fellowship of China Postdoctoral Science Foundation(2022TQ0178).

\ifCLASSOPTIONcaptionsoff
  \newpage
\fi



\bibliographystyle{IEEEtran}
\bibliography{IEEEabrv,bibi}

%

\begin{IEEEbiography}[{\includegraphics[width=1in,height=1.25in,clip,keepaspectratio]{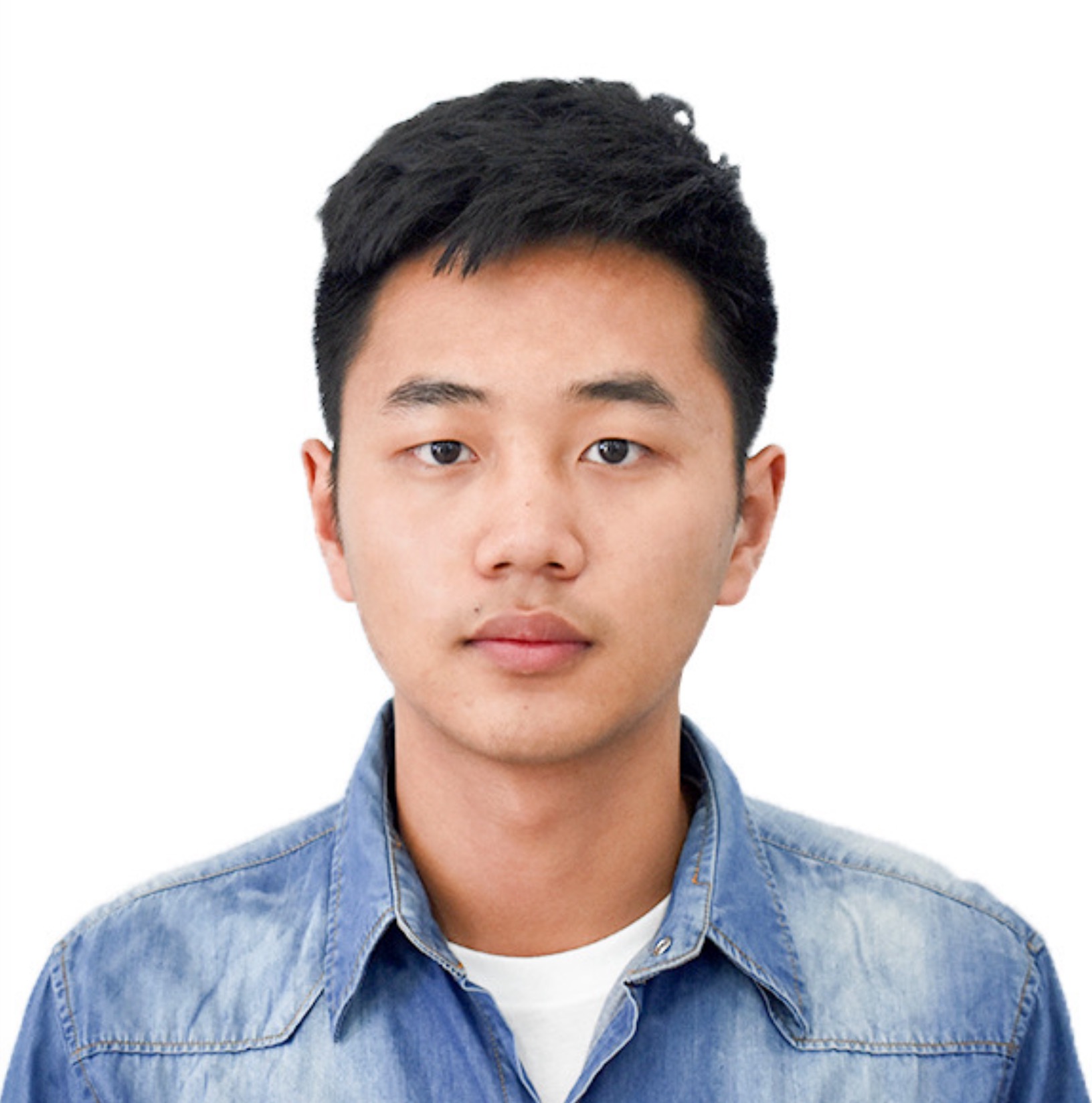}}]
{Peijie Sun} is a Postdoc at Tsinghua University. He received his Ph.D. degree from the Hefei University of Technology in 2022. He has published several papers in leading conferences and journals, including WWW, SIGIR, IEEE TKDE, IEEE Trans. on SMC: Systems, and ACM TOIS. His current research interests include data mining and recommender systems.
\end{IEEEbiography}

\begin{IEEEbiography}[{\includegraphics[width=1in,height=1.25in,clip,keepaspectratio]{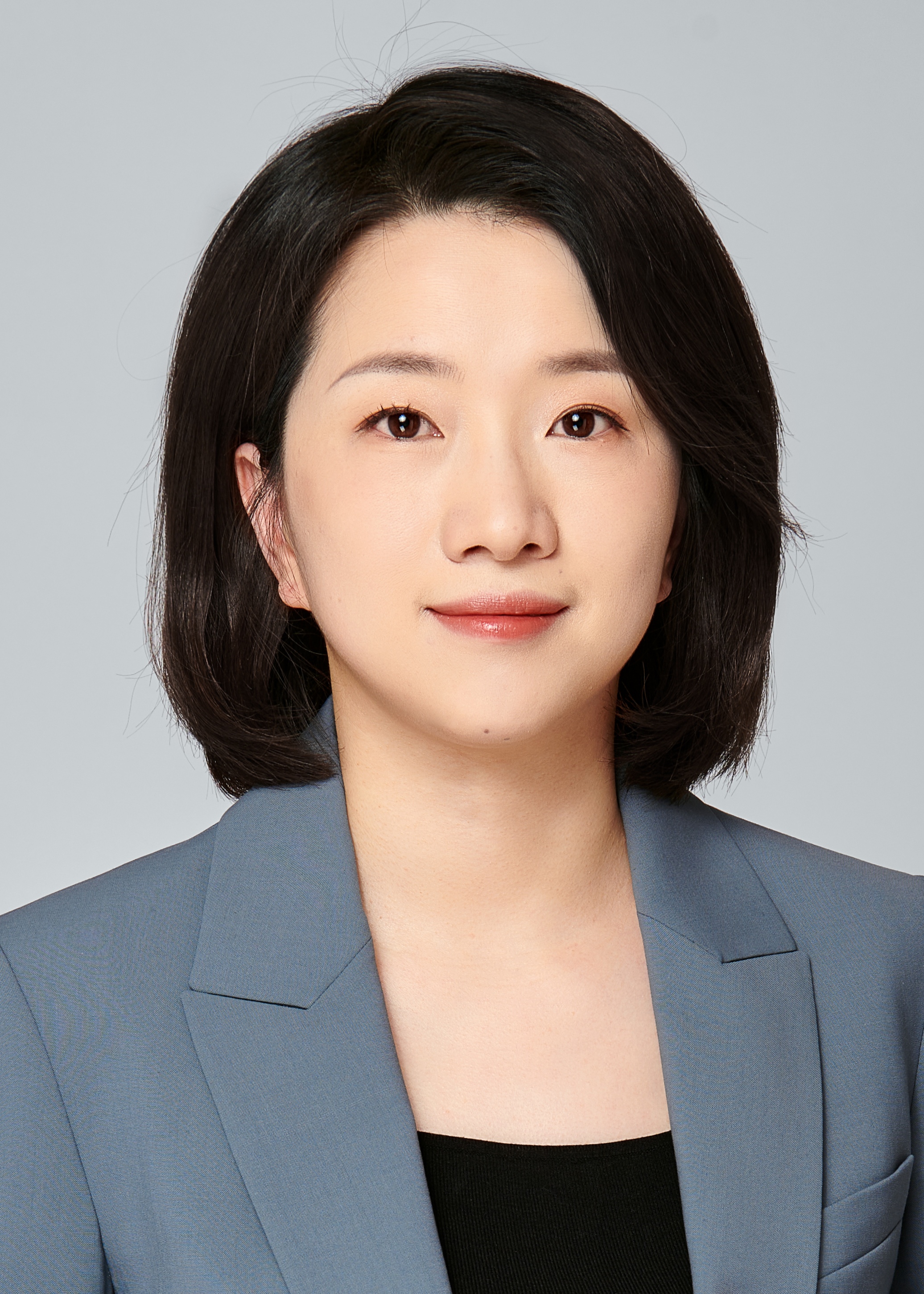}}]
{Le Wu (M'18)} is currently an associate professor at the Hefei University of Technology (HFUT), China. She received her Ph.D. degree from the University of Science and Technology of China (USTC). Her general areas of research interest are data mining, recommender systems, and social network analysis. She has published more than 40 papers in referred journals and conferences. Dr. Le Wu is the recipient of the Best of SDM 2015 Award, and the Distinguished Dissertation Award from the China Association for Artificial Intelligence (CAAI) 2017.
\end{IEEEbiography}

\begin{IEEEbiography}[{\includegraphics[width=1.0in,height=1.25in,clip,keepaspectratio]{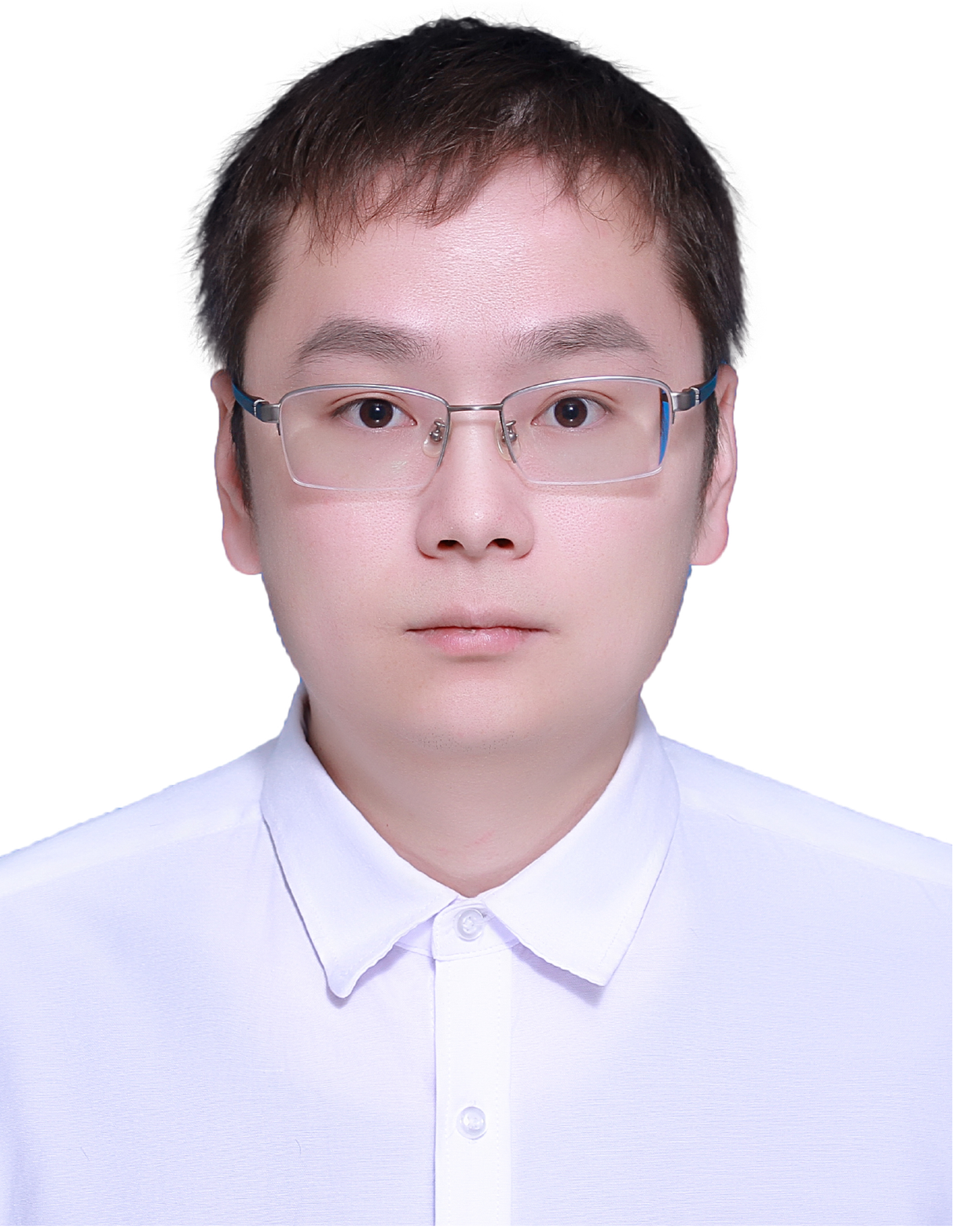}}]
{Kun Zhang} received a Ph.D. degree in computer science and technology from the University of Science and Technology of China, Hefei, China, in 2019. He is currently a faculty member at the Hefei University of Technology (HFUT), in China. His research interests include Natural Language Understanding and Recommendation Systems. He has published several papers in refereed journals and conferences, such as the IEEE TSMC:S, IEEE TKDE, ACM TKDD, AAAI, KDD, ACL, and ICDM. He received the KDD 2018 Best Student Paper Award.
\end{IEEEbiography}

\begin{IEEEbiography}[{\includegraphics[width=1in,height=1.25in,clip,keepaspectratio]{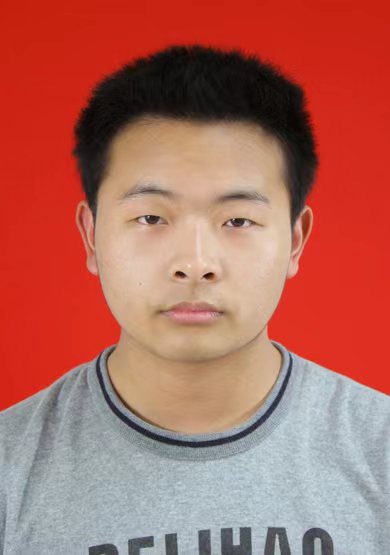}}]
{Xiangzhi Chen} is a Ph.D. student at Hefei University of Technology. His current research interests include data mining and recommender systems.
\end{IEEEbiography}


\begin{IEEEbiography}[{\includegraphics[width=1.0in,height=1.25in,clip,keepaspectratio]{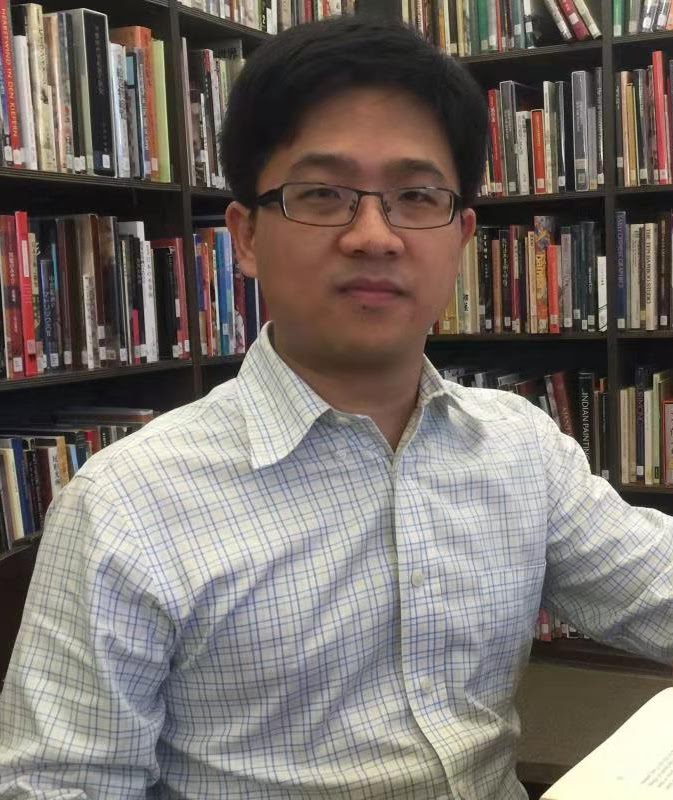}}]
{Meng Wang(SM'17)} is a professor at the Hefei University of Technology,
China. He received his B.E. degree and Ph.D. degree in the Special
Class for the Gifted Young and the Department of Electronic
Engineering and Information Science from the University of Science and
Technology of China (USTC), Hefei, China, in 2003 and 2008,
respectively. His current research interests include multimedia
content analysis, computer vision, and pattern recognition. He has
authored more than 200 book chapters, journal and conference papers in
these areas. He is the recipient of the ACM SIGMM Rising Star Award 2014.
He is an associate editor of IEEE Transactions on Knowledge and Data
Engineering (IEEE TKDE), IEEE Transactions on Circuits and Systems
for Video Technology (IEEE TCSVT), IEEE Transactions on Multimedia (IEEE TMM), and IEEE Transactions on Neural Networks and Learning Systems (IEEE TNNLS).
\end{IEEEbiography}




\end{document}